\documentclass[fleqn,usenatbib,useAMS]{mnras}

\usepackage{graphicx}	
\usepackage{amsmath}	
\usepackage{amssymb}	
\usepackage{multicol}        
\usepackage{bm}		
\usepackage{pdflscape}	
\usepackage{xspace}

\newcommand{\FINISH}{\bibliographystyle{mnras}\bibliography{example}\end{document}}

\newcommand{\GRS}{\mbox{GRS~1915+105}\xspace}
\newcommand{\RXTE}{\textit{RXTE}\xspace}
\newcommand{\MAXI}{\mbox{MAXI J1348--630}\xspace}

\newcommand{\obsnumber}{398\xspace}

\usepackage[T1]{fontenc}
\usepackage{ae,aecompl}

\usepackage{newtxtext,newtxmath}

\title[The corona of GRS 1915+105: A spectral-timing perspective]{The evolving properties of the corona of GRS 1915+105: A 
spectral-timing perspective through variable-Comptonisation modelling}

\author[Federico Garc\'ia et al.]{Federico Garc\'ia$^{1,2}$ 
\thanks{Contact e-mail: \href{mailto:fgarcia@iar.unlp.edu.ar}{fgarcia@iar.unlp.edu.ar}}, 
Konstantinos Karpouzas$^{1}$, 
Mariano M\'endez$^{1}$, 
Liang Zhang$^{3,4}$, 
Yuexin Zhang$^{1}$, 
\newauthor
Tomaso Belloni$^{5}$, 
Diego Altamirano$^{4}$
\newauthor
\\
$^{1}$  Kapteyn Astronomical Institute, University of Groningen, PO BOX 800, NL-9700 AV Groningen, the Netherlands \\
$^{2}$ Instituto Argentino de Radioastronom\'ia (CCT La Plata, CONICET; CICPBA; UNLP), C.C.5, (1894) Villa Elisa, Buenos Aires, Argentina\\
$^{3}$ Key Laboratory of Particle Astrophysics, Institute of High Energy Physics, Chinese Academy of Sciences, Beĳing 100049, People’s Republic of China\\
$^{4}$  School of Physics and Astronomy, University of Southampton Highfield Campus, Southampton SO17 1PS, UK\\
$^{5}$  INAF-Osservatorio Astronomico di Brera, via E. Bianchi 46, I-23807, Merate, Italy\\
}
\date{Accepted 2022 April 27. Received 2022 April 1; in original form 2022 February 4}

\pubyear{2022}

\begin{document}
\label{firstpage}
\pagerange{\pageref{firstpage}--\pageref{lastpage}}
\maketitle

\begin{abstract}
The inverse Compton process by which soft photons are up-scattered by hot electrons in a corona plays a fundamental role in shaping the X-ray spectra of black-hole (BH) low-mass X-ray binaries (LMXBs), particularly in the hard and hard-intermediate states. In these states, the power-density spectra of these sources typically show Type-C low-frequency quasi-periodic oscillations (QPOs). Although several models have been proposed to explain the dynamical origin of their frequency, only a few of those models predict the spectral-timing radiative properties of the QPOs. Here we study the physical and geometrical properties of the corona of the BH-LMXB GRS 1915+105 based on a large sample of observations available in the \RXTE archive. We use a recently-developed spectral-timing Comptonisation model to fit simultaneously the energy-dependent fractional rms amplitude and phase-lag spectra of the Type-C QPO in 398 observations. For this, we include spectral information gathered from fitting a Comptonisation model to the corresponding time-averaged spectra. We analyse the dependence of the physical and geometrical properties of the corona upon the QPO frequency and spectral state of the source, the latter characterised by the hardness ratio. We find consistent trends in the evolution of the corona size, temperature, and feedback (the fraction of the corona photons that impinge back onto the disc) that persist for roughly 15~years. By correlating our observations with simultaneous radio-monitoring of the source at 15~GHz, we propose a scenario in which the disc-corona interactions connect with the launching mechanism of the radio jet in this source.
\end{abstract}

\begin{keywords}
X-ray: binaries -- X-ray: individual (\GRS) -- accretion, accretion discs -- stars: black holes
\end{keywords}


\section{Introduction}

Black hole (BH) X-ray binary systems regularly show low-frequency (LF) quasi-periodic oscillations (QPOs) in the power density spectra (PDS) of their X-ray light curves \citep[see recent reviews on QPOs by][and references therein]{2016ASSL..440...61B,2020arXiv200108758I}. These narrow distinct peaks are usually characterised by their centroid frequency, $\nu$, and quality factor $Q = \nu / \Delta \nu$, where $\Delta \nu$ is their full-width half maximum (FWHM). According to these properties, and the shape and level 
of the underlying broad-band noise in the PDS, these LF QPOs were classified into three main types, namely A, B and C \citep{1999ApJ...526L..33W,2002ApJ...564..962R,2005ApJ...629..403C}. Among these three types, the type-C are the most common ones, characterised by variabilities of up to 20\% and narrow peaks usually with $Q \ga 10$. They are mainly observed in the Low-Hard and Hard-Intermediate states \citep{2000A&A...355..271B,2005Ap&SS.300..107H}, with frequencies from a few mHz to $\sim$10~Hz \citep[see][]{2012MNRAS.427..595M}, but also in the High-Soft and Ultraluminous states, reaching up to $\sim$30~Hz \citep{2000MNRAS.312..151R}. 

Several models had been proposed to explain the dynamical origin of type-C QPOs, based either on geometrical effects, associated to the Lense-Thirring precession frequency \citep[LTP,][]{1998ApJ...492L..59S,2006ApJ...642..420S,2009MNRAS.397L.101I}, or instabilities in the accretion flow \citep{1999A&A...349.1003T,2010MNRAS.404..738C}, but their physical origin remains a matter of debate. The geometrical scenario proposed by \cite{2009MNRAS.397L.101I} based on relativistic precession requires two main components: a cool optically thick and geometrically thin accretion disc \citep{1973A&A....24..337S} which is truncated at $r_t > r_{\rm ISCO}$, where $r_{\rm ISCO}$ is the radius of the innermost-stable circular orbit, and a hot, geometrically thick, accretion flow inside $r_t$ \citep{1997ApJ...489..865E,1997MNRAS.292L..21P}. In this model, the type-C QPOs are produced by the extended hot inner flow that modulates the X-ray flux as it precesses at the LTP frequency \citep{2007ApJ...668..417F}. The model predicts that this effect should increase with source inclination. In this framework, the broad-band noise in the PDS, instead, would arise from variations in the mass accretion rate that propagate from the outer regions of the disc towards the BH \citep{2013MNRAS.434.1476I}. In general, the frequency of the type-C QPO increases as the source moves from harder to softer states. In the LTP model \citep{2009MNRAS.397L.101I}, the change in hardness is interpreted as a variation in the outer radius of the hot inner flow. In models involving an extended Comptonising medium, or corona \citep{1975ApJ...195L.101T,1979Natur.279..506S}, the hardness changes are attributed to variations in the physical properties of this medium, like the optical depth or electron temperature, or in geometrical properties, like its size \citep{1997ApJ...480..735K}.

Despite the big efforts pursued to understand the physical origin of the QPO dynamics, less attention has been put to quantitatively explain the time-dependent radiative properties of these QPOs, which are given by the energy and frequency dependence of the QPO amplitudes and phase lags. \cite{2006MNRAS.370..405S} analysed the variability-amplitude (RMS) spectra of the Type-C QPO in several BH XRBs. They found that even in observations where a soft disc-like thermal component was evident in the time-averaged spectra, this component was not found in the RMS spectrum of the QPOs, indicating that the QPO strongly modulates the Comptonised emission, although not the emission coming directly from the disc. In this sense, the lack of a disc imprint in the QPO spectrum is a complication for QPO models based on disc oscillations \citep{2020arXiv200108758I}.

Some models for the QPO phase lags can be found in  \citep{1998MNRAS.299..479L,2012ApJ...752L..25S,2013ApJ...779...71M}. The successive inverse Compton up-scattering of soft-photons by hot electrons was suggested as the physical explanation for the hard lags observed in the broad-band noise component in the PDS of Cyg~X-1 \citep{1988Natur.336..450M}. In turn, broad-band soft lags can be naturally produced by reverberation off the accretion disc. That is the case in a recently-discovered BH candidate (MAXI~J1820+070), where a small contracting corona of a few gravitational radii ($R_g = GM/c^2$, were $G$ is the gravitational constant, $M$ is the mass of the compact object and $c$ the speed of light) was suggested based on the frequency evolution of the soft reverberation lags measured in the source \citep{2019Natur.565..198K}. However, the lags associated to the LF QPOs are usually much larger than those found in the broad-band noise, and cannot be explained by reverberation from a small-scale hot inner flow \citep{2015ApJ...814...50D}. 

In a Comptonisation scheme, soft lags can be generated through a feedback process, if a significant fraction of the up-scattered photons impinge back onto the soft-photon source \citep{2001ApJ...549L.229L}. A variety of models \citep[e.g.,][]{2000ApJ...538L.137N,2009MNRAS.397L.101I,2016MNRAS.461.1967I} invoking different physical processes can explain the QPO phase lags, but not many of those models can predict the energy-dependent rms amplitude of the variability. \cite{2015MNRAS.447.2059M} suggested that the rms amplitude of the type-C QPO is higher for high inclination sources compared to low ones. More recently, \cite{2017MNRAS.464.2643V} found evidence that in low-inclination sources, above a certain QPO frequency, the phase-lags of the QPO are hard and increase with QPO frequency, meanwhile, for high-inclination sources, the opposite is found: QPO phase-lags are soft and decrease as the QPO frequency increases. These inclination dependence may favour a geometrical origin, but a full explanation for the very diverse observational findings in the subject is still lacking.

\GRS is a very particular BH LMXB. It has been in outburst since its discovery \citep{1992IAUC.5590....2C,1994ApJS...92..469C} and did not show the typical Q-shape in the HID. Being active during the whole 15-years lifetime of the {\it Rossi} X-ray Timing Explorer \citep[\RXTE,][]{1993A&AS...97..355B} mission, \GRS is hence one of the most comprehensively observed and studied sources in the \RXTE archive \citep[for a review on \GRS, we refer the reader to][]{2004ARA&A..42..317F}. \GRS was the first XRB to show superluminal ejections of synchrotron-emitting components in the radio band, through a relativistic jet \citep{1994Natur.371...46M}. Using Very-Long Baseline Array observations, \cite{2014ApJ...796....2R} measured the trigonometric parallax to \GRS yielding a distance estimate of 8.6$^{+2.0}_{-1.6}$~kpc and a BH mass of 12.4$^{+2.0}_{-1.8}$~M$_\odot$. 

The spectral-timing properties of the LF QPOs in \GRS had been studied in several papers \citep{1999ApJ...513L..37M,2003A&A...397..729V,2004ApJ...615..416R}. Remarkably, a change in the sign of the phase lags between soft and hard X-rays in the type-C QPO has been recognised at QPO frequencies around 2~Hz \citep{2000ApJ...541..883R,2013ApJ...778..136P}. This effect has been thoroughly studied in a recent work by  \cite{2020MNRAS.494.1375Z}. Analysing the energy-dependent lag spectra of $\sim$600 observations of the type-C QPO, the authors find that the slope of the lag spectra has a clear dependence on the QPO frequency, being positive below $\sim$1.8~Hz, flat or zero at frequencies around $\sim$2~Hz, and becoming negative at higher QPO frequencies. At the same time, \cite{2020MNRAS.494.1375Z} show that the fractional-variability (rms) amplitude of the QPO is maximum when the lags are flat (around $\sim$2~Hz). 

Motivated by the recent model from \cite{2020MNRAS.492.1399K} and the unique extensive data set available for \GRS, in this paper we present a systematic study of the evolving properties of the corona of \GRS by fitting a variable-Comptonisation model to the energy-dependent phase lag and fractional rms amplitude spectra of the Type-C QPO in \obsnumber \RXTE observations. To do this: i) we measure the full set of energy-dependent rms-amplitudes of the Type-C QPOs detected by \RXTE, ii) we take the energy-dependent phase-lags measurements available in \cite{2020MNRAS.494.1375Z}, and iii) we also use the time-averaged physical properties of the corona of \GRS fitted to the same set of observations in \cite{2022mariano}. In Sec. 2 we present our sample of observations and the methods employed to measure the QPO variability and the time-averaged spectral properties. In Sec. 3 we briefly describe the spectral-timing Comptonisation model used. In Sec. 4 we show the results obtained with the spectral-timing fits to the data and we discuss them in Sec.~5.

\section{Observations and data analysis}
\label{sec:data_analysis}

\begin{figure*}
 \includegraphics[width=\columnwidth]{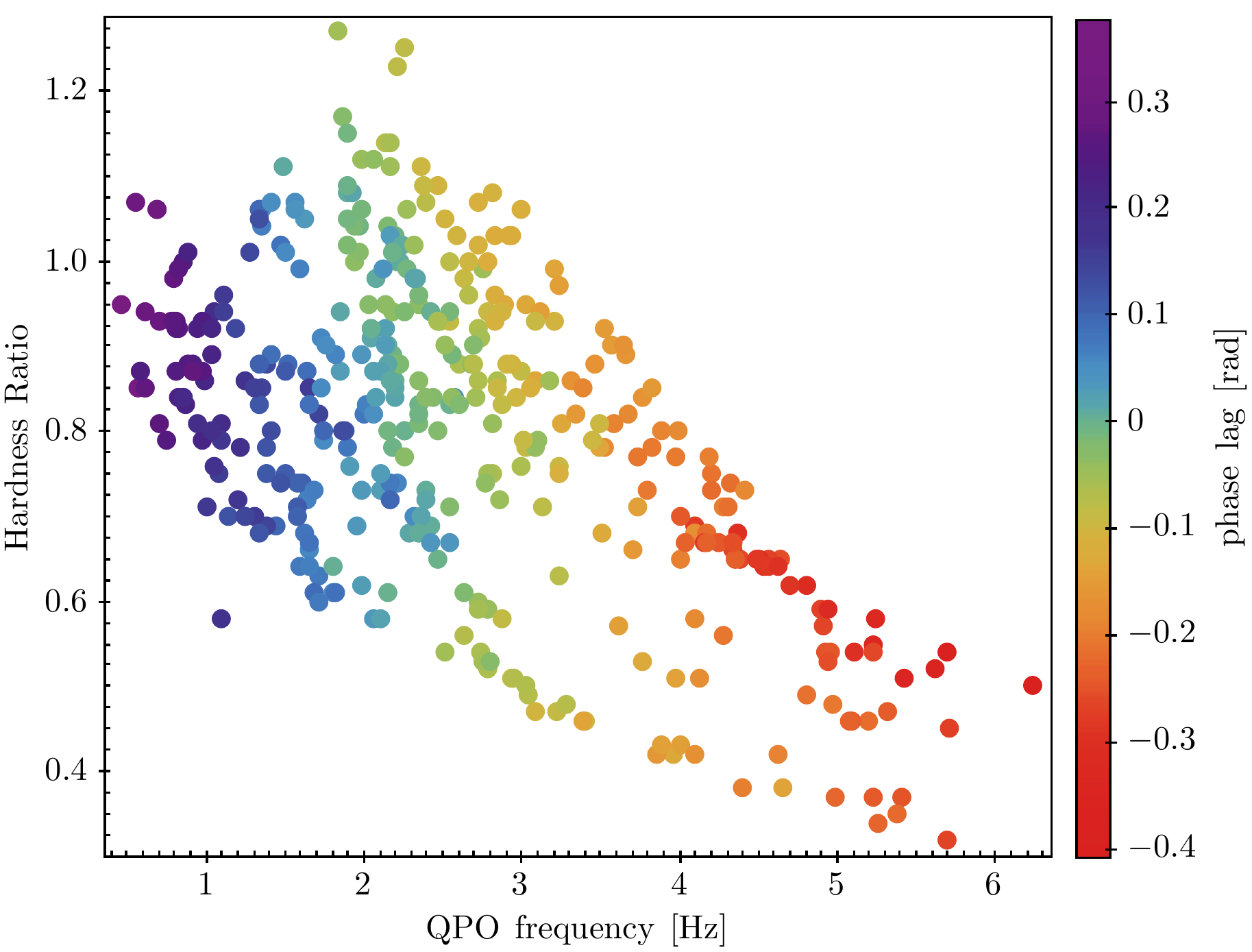}\,\,\,\,
 \includegraphics[width=\columnwidth]{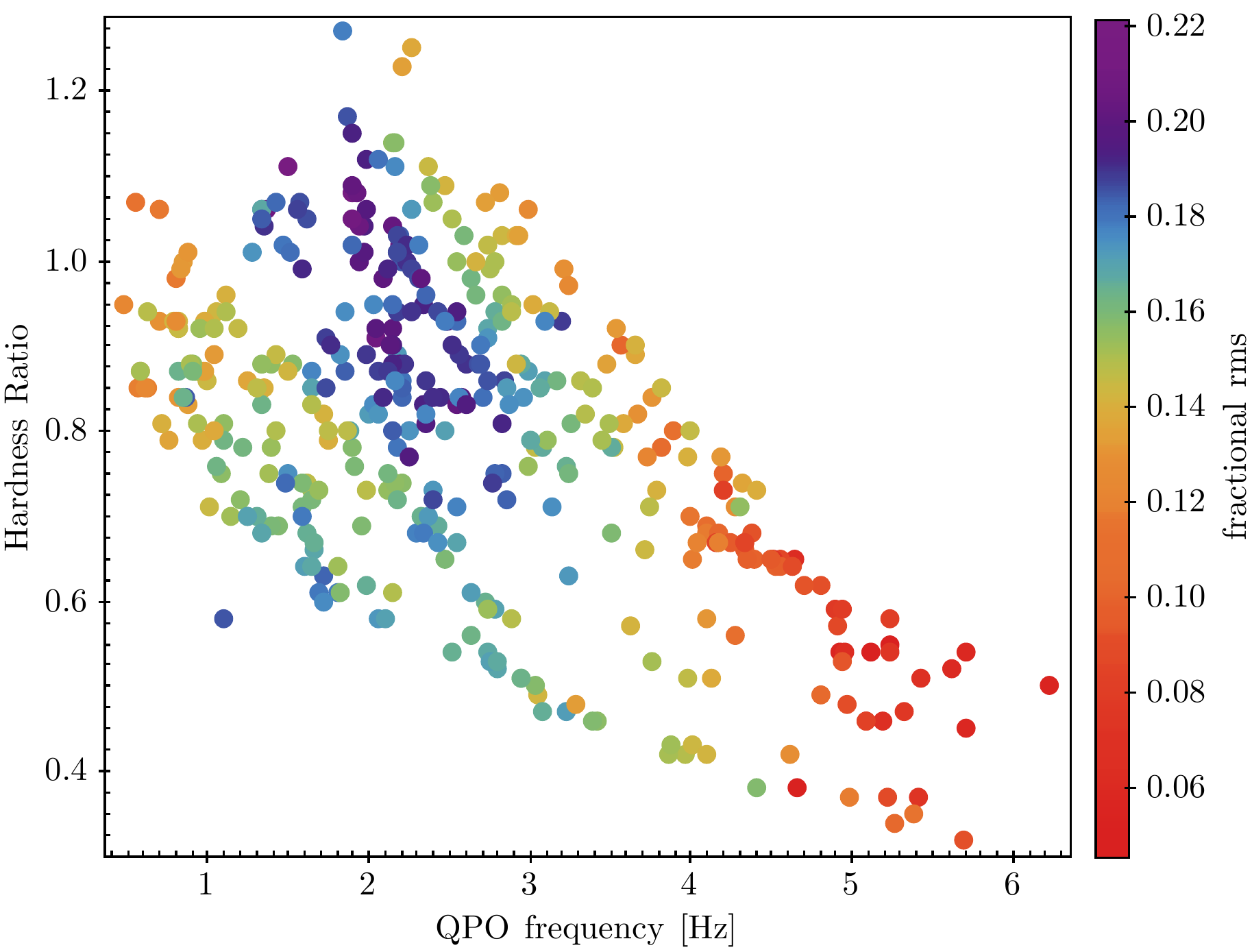}
 \caption{Phase lags and fractional rms (in the full PCA energy range) of the Type-C QPO as a function of QPO frequency and hardness ratio (in Crab units). Average errors of the QPO frequency and hardness ratio are $\pm$0.05~Hz and $\pm$0.001~Crab units, respectively (smaller than the size of the points in the plot). Average errors in phase lag and fractional rms are 0.01 rad and 0.003, respectively.}
 \label{fig:lags_and_rms}
\end{figure*}

\cite{2020MNRAS.494.1375Z} examined the full set of observations of \GRS obtained with the Proportional Counter Array (PCA) on-board the {\em Rossi} X-ray Timing Explorer \citep[\RXTE;][]{1993SPIE.2006..324Z} between 1996 and 2012. For each observation, \cite{2020MNRAS.494.1375Z} calculated a Fourier power density spectrum (PDS) in the full energy range (\textsc{PCA} channels 0 to 249) every 128~s, with 1/128~s time resolution (which corresponds to a Nyquist frequency of 64~Hz). They subsequently averaged the 128-s power spectra within each observation, subtracted the Poisson contribution, and re-normalised them to units of fractional rms squared per Hz. Since the source count rate was always very high, they ignored the background contribution in the conversion to rms units. Finally they re-binned the power spectra using a logarithmic step in frequency, such that the size of each frequency bin is exp(1/100) times larger than the previous one. In order to fit the resulting power spectra, the authors constructed a model \textsc{XSPEC v12.9} \citep{1996ASPC..101...17A} consisting of a sum of Lorentzian functions to represent the broad-band noise component and the different QPOs \citep[see][for more details of the analysis]{2020MNRAS.494.1375Z}.

Following \cite{2020MNRAS.494.1375Z}, in this paper we selected the observations that have a significant QPO consistent with the characteristics of the type-C QPOs, namely: the observations are in the $\chi$ state, equivalent to the HIMS, the power spectrum showed at least one narrow peak ($Q\approx5-15$, and strictly $Q>4$) significantly detected (>3$\sigma$) on top of the broad-band noise component. As in \cite{2020MNRAS.494.1375Z}, we excluded observations in which the QPO frequency changed significantly during one observation. We then cross-matched our \RXTE sample with publicly-available observations of \GRS with the Ryle radio telescope at 15~GHz \citep{1997MNRAS.292..925P}, considering that an X-ray and radio observation where ``simultaneous'' if they were performed within 2~days. This process yields a final data set consisting of 410 observations, with the type-C QPO detected within the 0.4--6.3~Hz frequency range \citep{2022mariano}.

\subsection{Spectral-timing analysis of the low-frequency Type-C QPO}
\label{sec:rmslags}

Following the method described in \cite{1997ApJ...474L..43V} and \cite{1999ApJ...510..874N}, \cite{2020MNRAS.494.1375Z} produced frequency-dependent phase-lags between the 2--5.7 and 5.7--15~keV energy bands for each observation. Since the \RXTE data set contains observations performed during different PCA calibration epochs (3--5), the authors actually selected the closest absolute channels matching these energy bands considering the PCA channel-to-energy gain factor\footnote{\url{https://heasarc.gsfc.nasa.gov/docs/xte/e-c_table.html}}. The QPO phase lags were obtained by averaging the lag-frequency spectrum around the QPO centroid, $\nu_0 \pm FWHM/2$, where $FWHM$ is the full-width at half-maximum of the Lorentzian used to fit the QPO profile.
Furthermore, following \cite{2014A&ARv..22...72U}, for each observation in the sample, \cite{2020MNRAS.494.1375Z} also calculated energy-dependent phase lags at the QPO, using the 4--6 keV, 6--8 keV, 8--11 keV, 11--15 keV, 15--21 keV, and 21--44~keV energy bands as subject bands, and the 2--4~keV energy band as the (soft) reference band. \citep[We refer the reader to][for more details of the spectral-timing analysis]{2020MNRAS.494.1375Z}. In our paper, phase lags are positive when they are hard, meaning that the hard photons lag the soft ones.

In order to measure the energy-dependent fractional-rms spectra, we proceeded in a slightly different way. Since we were also interested in studying the full frequency range where the PDS shows significant variability, we used clean event files with a time resolution of 1/512~s, which correspond to a Nyquist frequency of 256~Hz, divided into segments of 16~s. We generated the averaged PDS, subtracted the Poisson level, and applied a logarithmic rebin in frequency (using a factor $\exp(1/100)$), considering the full set of PCA channels. We also split the light-curves into 5 subsets using channels 0 to 13, 14 to 35, 36 to $\sim$50, $\sim$51 to $\sim$103, $\sim$104 to 249, respectively, to obtain the energy-dependent rms, which in this case has a slightly lower energy resolution than in the case of the lags (as we used the data modes with higher time resolution). Here the ``$\sim$'' symbol indicates that we use the closest channel boundary available in each observation. In some observations, performed in single-bit mode, the channels start from channel 8, and hence we use 8--13 as our first subset of channels. To define more or less the same energy bands, we calculate the energy channels independently for each observation, based on the channel-to-energy conversion of the epoch when each observation was performed. To properly estimate the fractional rms and its error bar, we take into account the background rate, which becomes a significant fraction of the total observed count rate at the hardest X-ray bands. We notice that the PDS of channels $\sim$104 to 249 is always noisy, and the QPO is not detected, and thus we exclude this channel range in our analysis. Following the method described above, we use the best-fitting model of the full PDS to fit the PDS obtained for each energy band. We obtain the rms of the QPO as the square root of the normalisation, $N$, of the Lorentzian that fits the QPO. Taking into account the propagation of errors from the count rates of the source, $S$, and the source plus background, $SB$, on the rms, we obtain its uncertainty, $\sigma_{\rm rms}$, from:

\begin{equation}
    \sigma_{\rm rms} = {\rm rms} \sqrt{\frac{1}{4}\left(\frac{\sigma_N}{N}\right)^2 + \frac{1}{4}\left(\frac{\sigma_{\textit{SB}}}{\textit{SB}}\right)^2 + \left(\frac{\sigma_S}{S}\right)^2}, 
\end{equation}
where $\sigma$ denotes the corresponding variance of each quantity. In this process we found 12 observations which had energy channels where the QPO was not significant, and thus we decided to exclude them from the subsequent analyses. These led us with a final data set consisting of \obsnumber observations for which we have: time-averaged spectral data, energy-dependent phase-lags and fractional rms amplitudes of the Type-C QPO, as well as radio-flux measurements at 15~GHz.

\subsection{Hardness ratio}
\label{sec:hr}

For each individual \RXTE observation of \GRS analysed in this paper, we calculate a hardness ratio ($HR$) between the background-subtracted count rates of the source in the 13--60~keV, {\em hard}, and the 2--7~keV, {\em soft}, band, using the closest absolute channels matching these energy ranges according to the PCA gain epoch. We correct the observed count rates for instrumental dead time and normalise them to the rate of the Crab in the same bands beforehand 
\citep[see e.g.,][]{2008ApJ...685..436A}. 

In Figure~\ref{fig:lags_and_rms} the colours of the points give the QPO phase lags (left panel) and fractional rms amplitude (right panel) as a function of $HR$ (y-axis) and QPO frequency ($\nu_0$, x-axis) for the \obsnumber observations of \GRS, in what we, from now on, call $QPO-HR$ diagrams \citep[see also][]{2001ApJ...558..276T,2022mariano}. On the one hand, the phase lags show a strong dependence on the QPO frequency \citep[as previously discussed by ][]{2020MNRAS.494.1375Z} and a weak dependence on $HR$ (the colour gradients are nearly horizontal in the plot). On the other hand, the rms amplitudes show a more complex dependence on the $QPO-HR$ diagram. Observations with the highest rms values cluster around 1.5--2.5~Hz, but in observations with a harder spectrum ($HR>0.8$) the rms amplitude of the QPO is significantly higher than in observations with a softer spectrum ($HR<0.8$). The lowest rms values are found for QPO frequencies either $\la$1~Hz or $\ga$4~Hz, independently of whether the lags are either  soft or hard \citep[see][for more details]{2020MNRAS.494.1375Z}.

\subsection{Time-averaged spectra}
\label{sec:timeavg}

\begin{figure*}
 \includegraphics[width=\columnwidth]{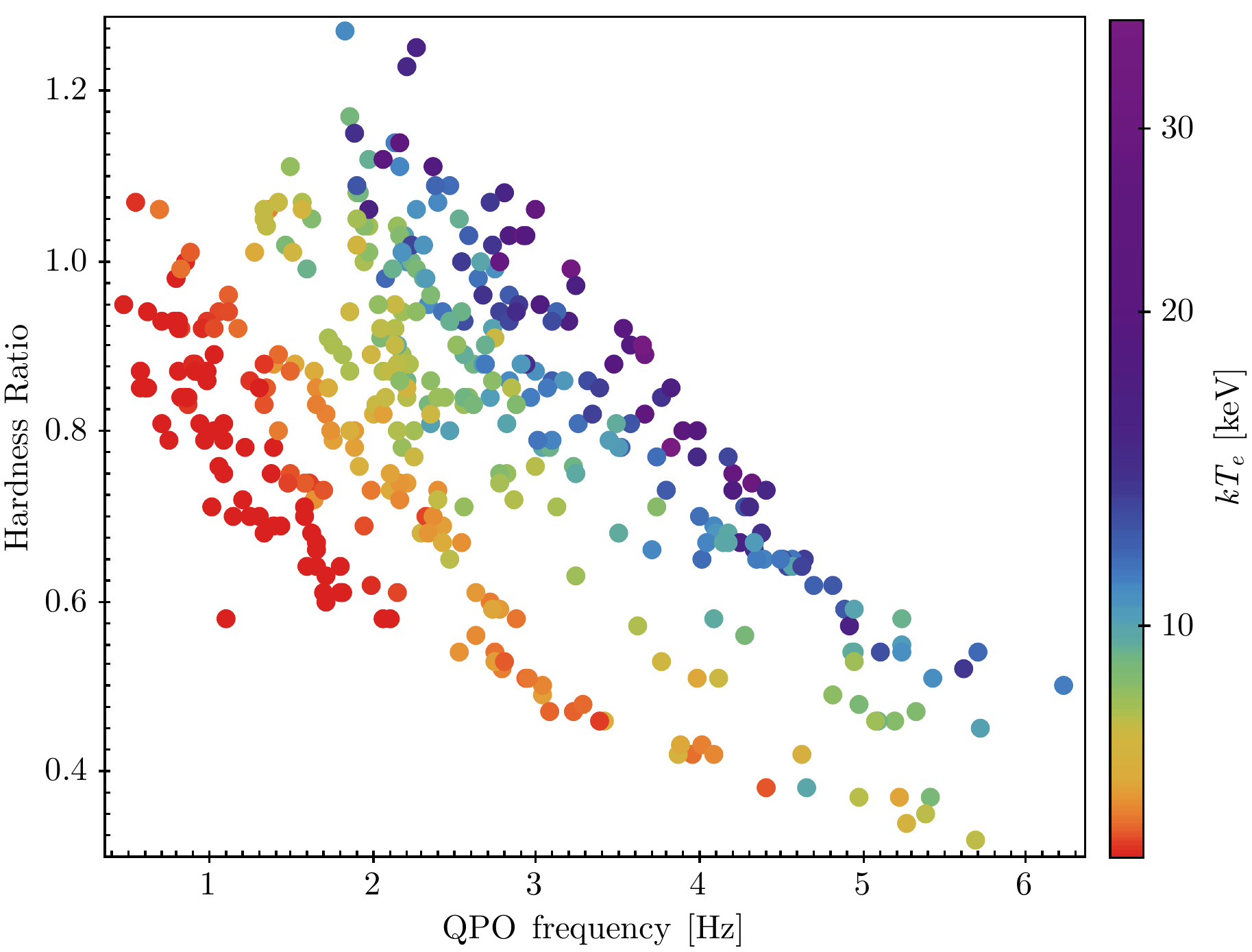}\,\,\,\,
 \includegraphics[width=\columnwidth]{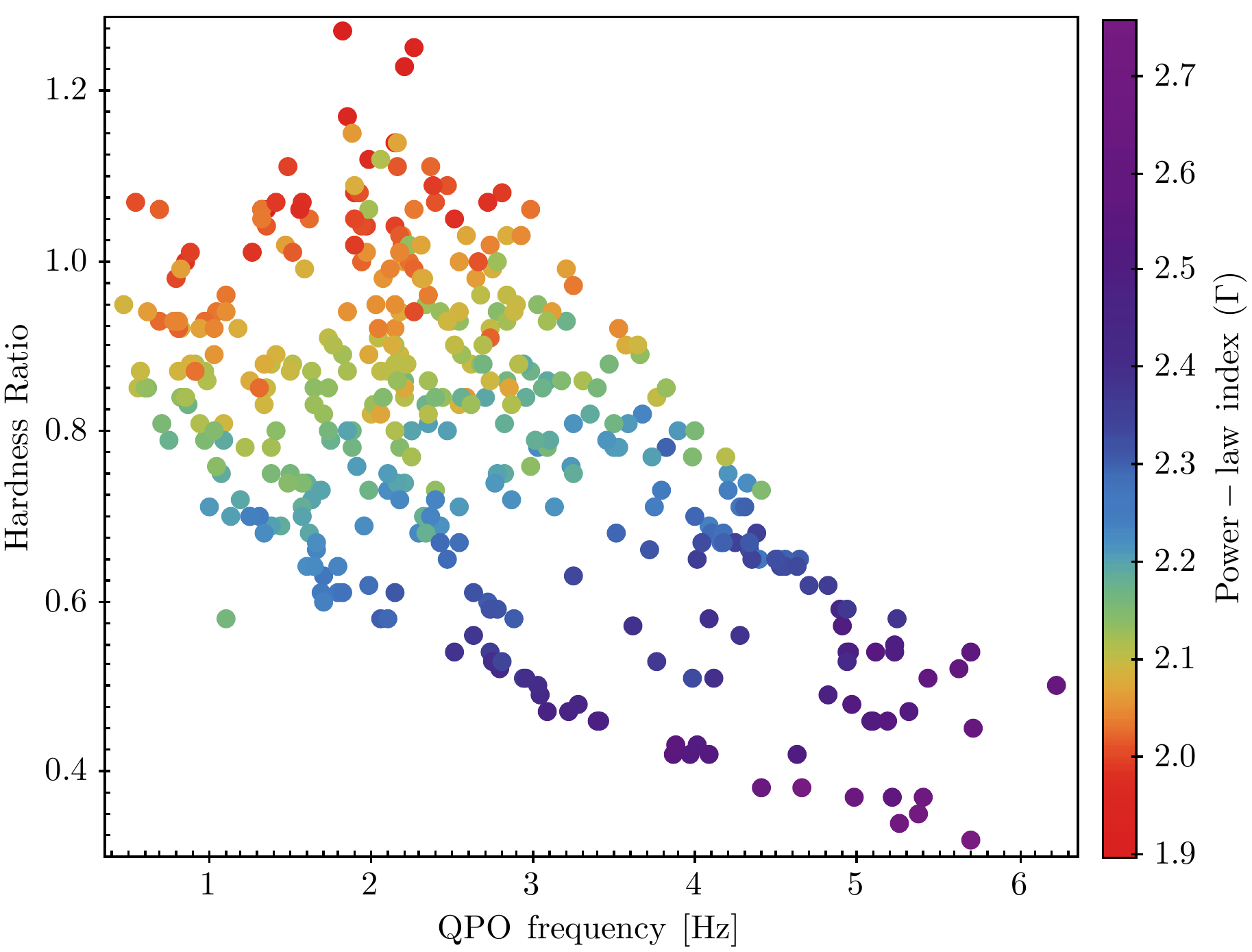}
 \caption{Electron temperature ($kT_e$) and power-law index ($\Gamma$) of the corona as a function of QPO frequency and Hardness Ratio \citep[see also][]{2022mariano}. The best-fitting values are obtained fitting a Comptonisation model to the time-averaged spectra. Average errors in $kT_e$ and $\Gamma$ are $\pm$0.75~keV and $\pm$0.14, respectively.}
 \label{fig:kTe_and_Gamma}
\end{figure*}

In this Section we briefly describe the methods followed by \cite{2022mariano} to obtain, analyse and fit the time-averaged spectra of the \obsnumber observations considered in this paper. For each observation in the sample, \cite{2022mariano} extracted dead-time corrected energy spectra using the \RXTE Standard 2 data. They used \textsc{pcabackest} and \textsc{pcarsp} available in \textsc{headas v6.27} to extract background spectra and produce response files, respectively, and corrected the energy spectra for dead time. Using the model \textsc{vphabs*(diskbb+gauss+nthcomp)} in \textsc{XSPEC v12.9} \citep{1996ASPC..101...17A}, \cite{2022mariano} performed a joint fit of the full set of energy spectra, linking the hydrogen column density, $N_{\rm H}$ in the \textsc{vphabs} component to account for the interstellar absorption along the line of sight to the source, using the chemical abundances and cross sections given by \cite{2000ApJ...542..914W} and \cite{1996ApJ...465..487V}, respectively. Since $N_{\rm H} \approx 6\times10^{22}$~atoms~cm$^{2}$ is quite high in the direction of \GRS, they also let free to vary the Fe abundance of this absorption component to account for an Fe absorption edge at $E \sim 7.1$~keV that was apparent in the fitting residuals. The \textsc{diskbb} component stands for the emission from an geometrically-thin and optically-thick accretion disc \citep{1973A&A....24..337S}, with a temperature $kT_{\rm dbb}$ at its inner radius. The \textsc{nthcomp} component represents the inverse-Compton emission from the corona \citep{1996MNRAS.283..193Z, 1999MNRAS.309..561Z}, and is parametrized by the temperature of the source of soft photons, $kT_{\rm bb}$, that are up-scattered in the corona of hot electrons with temperature $kT_e$, and a power-law index, $\Gamma$. In the fits, the authors assumed that the seed-photons source was the accretion disc, and thus linked the $kT_{\rm bb}$ parameter to $kT_{\rm dbb}$, separately for each observation. In this inverse-Compton model the optical depth, $\tau$, of the corona (assumed to be homogeneous) is a function of the electron temperature, $kT_e$, and the power-law index, $\Gamma$:
\begin{equation}
\tau(kT_e, \Gamma) = \sqrt{ 2.25 + \frac{3}{\frac{kT_e}{m_e c^2}\left[(\Gamma+0.5)^2-2.25\right]}}, 
\label{eq:gamma}
\end{equation}
where $m_e$ is the rest mass of the electron and $c$ is the speed of light. Finally, the \textsc{gauss} component was used to represent a broad Fe emission line at $\sim 6.4-7$~keV, which arises due to reflection of corona photons off the accretion disc \citep{2009Natur.459..540F}. 

By combining the data presented in the previous Sections, in Figure~\ref{fig:kTe_and_Gamma} we show the dependence of $kT_e$ (left panel) and $\Gamma$ (right panel) upon the $HR$ and QPO frequency ($QPO-HR$ diagrams). Remarkably, $kT_e$ shows a very smooth diagonal gradient in the diagram, with values spanning from low temperatures ($\la$5--7~keV, red colour) at the bottom-left side of the plot to high temperatures (15--40~keV, blue colour) at the top right \citep{2022mariano}. Moreover, the power-law index of the Comptonisation component, $\Gamma$, also shows a smooth gradient in the $QPO-HR$ diagram. However, in this case the gradient is vertical, being strictly dominated by the $HR$ and quite independent of the QPO frequency. As expected, the $HR$ works as a proxy to the power-law index when the spectra is mainly driven by the Comptonisation component.

\section{The spectral-timing Comptonisation model}
\label{sec:model}

In this paper, we extensively use the variable-Comptonisation model from \cite{2020MNRAS.492.1399K} and \cite{2021MNRAS.503.5522K}. This model is a numerically-efficient implementation and extension of the model early-proposed by \cite{1998MNRAS.299..479L} and \cite{2001ApJ...549L.229L}, and more recently by \cite{2014MNRAS.445.2818K}. Given that the fractional rms amplitude of different types of QPOs in several sources increases with energy \citep[e.g.,][and references therein]{2006MNRAS.370..405S,2013MNRAS.435.2132M}, this family of models were built to explain the radiative properties of any QPO as an oscillation of the time-averaged, or steady-state, spectrum, in terms of the inverse-Compton process. The main idea is that the observed energy spectrum at the QPO frequency, both its energy dependent variability amplitude (the fractional rms) and phase lags, arises due to the coupled oscillations of the physical properties of the system, such as the temperature of the corona and the temperature of the source of seed photons. Hence, by fitting the spectral-timing properties of the QPO, the model provides physical properties of the corona that are otherwise not directly accessible through fits to the full time-averaged spectrum. This model does not explain the dynamical origin of the QPO, and assumes that the QPO frequency can be produced by any of the models previously proposed for that \citep[see e.g.,][]{2009MNRAS.397L.101I}.

The model of \cite{2020MNRAS.492.1399K} considers a spherical blackbody with temperature $kT_s$ as the source of seed photons, which is enshrouded by corona of hot electrons with temperature $kT_e$; the corona is a spherically-symmetric shell of size, $L$, with constant density and optical depth, $\tau$. Feedback onto the soft-photon source of up-scattered photons in the corona is also incorporated in the model, and controlled by the so-called feedback-fraction parameter, $0 \le \eta \le 1$, which is defined as the fraction of the flux of the seed-photon source which arises after the feedback process. In the inverse-Compton process, the soft photons from the seed source gain energy from the hot electrons in the corona. When the electrons give energy to the photons, in turn, they cool down. Notwithstanding, astrophysical coronas are long-lived, and thus there must be a source that provides energy to the corona at a rate $H_{\rm ext}$, the external heating rate, that keeps the corona in thermal equilibrium. 

The steady state of the model, which is also the expected time-averaged spectrum,  coincides with the \textsc{nthcomp} model and is the numerical solution of the stationary Kompaneets equation  \citep{1957JETP....4..730K}. In order to obtain the variability in the QPO, the model numerically solves a linearised version of the time-dependent Kompaneets equation, assuming that $kT_e$, $kT_s$, and $H_{\rm ext}$ undergo oscillations with small relative amplitudes $\delta kT_e$, $\delta kT_s$ and $\delta H_{\rm ext}$, respectively, at the QPO frequency. Those oscillations reflect into complex variability amplitudes of the spectrum that can be compared to the energy-dependent fractional rms amplitude and phase lags at the QPO frequency.

In this model, the feedback parameter, $\eta$, has a strong impact on the sign of the slope of the lag spectrum. On the one hand, Comptonisation naturally produces hard or positive lags \citep{1988Natur.336..450M} as, on average, photons need more interactions, and thus longer times inside the corona, to reach higher energies. On the other hand, if a significant portion of the Comptonised photons impinge back onto the soft-photon source ($\eta \gg 0$) the process leads to soft or negative lags \citep{2001ApJ...549L.229L}. In the same way, in the Comptonisation model, increasing both $L$ or $\tau$ leads to increasing slopes in the lag-energy spectra (either for positive or negative lags), since photons experience more scatterings before leaving the electron cloud. In turn, $\delta H_{\rm ext}$ has no effect on the shape of variability spectrum (either lags or rms). This parameter acts as a normalisation of the fractional rms spectrum: the larger $\delta H_{\rm ext}$, the larger the fractional variability amplitude. 

\cite{2020MNRAS.492.1399K} demonstrated that the model can successfully fit the energy-dependent fractional-rms amplitude and time lags of the lower kilohertz QPO of the neutron-star LMXB 4U~1636--53. 
Recently, \cite{2021MNRAS.501.3173G} used this model to successfully explain the spectral-timing properties of the low-frequency Type-B QPO in \MAXI. 
Furthermore, using a limited sample of nine observations, \cite{2021MNRAS.503.5522K} showed that the model can also explain the spectral-timing properties of the Type-C QPO in \GRS. In this paper we apply the same model to fit the full set of \obsnumber observations previously described. 
For this, we incorporate information gathered from the time-averaged spectra by fixing the $kT_e$ and $\Gamma$ parameters of the spectral-timing model to the best-fitting values of the time-averaged energy spectrum of each observation, best-fitted with an \textsc{nthcomp} model as explained above in Sec.~\ref{sec:timeavg}. We opt not to take the $kT_s$ from the time-averaged data, given that the soft-photon spectrum in our spectral-timing model is based on a black-body spectrum, while the time-averaged spectra were fitted using a multi-colour disc blackbody model. By doing this, we end up with four free physical parameters to fit to the QPO variability, namely the temperature of the soft-photon source, $kT_s$, the corona size, $L$, the feedback fraction, $\eta$, and the amplitude of the external heating rate, $\delta H_{\rm ext}$. We also fit an additive constant to the phase-lag spectra to take into account the reference-lag angle (which is physically meaningless). 

For each observation, we simultaneously fit the lag spectrum in the full $\sim$2--41~keV energy range and the fractional-rms spectrum up to $\sim$20~keV. We discard the highest-energy channel of the rms spectrum, as the spectral-timing model systematically predicts larger rms amplitudes than observed, most likely due to the absence of a reflection component in the model \citep[for more details, we refer to the model paper,][]{2020MNRAS.492.1399K}. In all the paper we will assume that the seed-photon source is a sphere with a radius of 250~km, consistent with the typical inner radius of the accretion disc at $\sim$2~Hz for a $\sim$12~M$_\odot$ BH like the one in \GRS \citep{2014ApJ...796....2R}, considering the LTP frequency. We note that the shape of the model is rather insensitive to this parameter (see Appendix~\ref{app:linearscheme}), varying by a factor $\la$1.05 within the range set by the estimated inner radius of the accretion disk \citep[$\gtrsim$50~km,][]{2022mariano} and the maximum LTP radius, corresponding to the minimum QPO frequency observed ($\lesssim$500~km).

In order to fit such a large data set consisting of \obsnumber observations, and considering the relatively low speed of the fitting process due to the numerical characteristics of the model, which involve non-trivial matrix inversions, we create a set of 17 multi-dimensional grids of pre-calculated models ({\em table models}), for 10 QPO frequencies in the 0.25--2.5~Hz range and 7 in the 3.0--6.0~Hz range. We construct each of these table models using either uniform or logarithmic grids for each free physical parameter, covering the range of values indicated in Table~\ref{tab:table}. Those values were chosen guided by both the best-fitting time-averaged spectral parameters from \cite{2022mariano} and those from the spectral-timing analysis of \cite{2021MNRAS.503.5522K}. The resulting table models consist of 315\,000~rows of 144 logarithmic energy bands covering the $\sim$0.5--50~keV energy range. Once the table models are created, for each observation, we take the closest-in-frequency table model available and we load it into \textsc{XSPEC} to simultaneously fit the fractional rms and phase lags. We firstly explore the full table model using the \textsc{steppar} function to compute the best-fitting model at each row of the table. Once we identify the row of the table that minimises the residuals of the fit, we run the \textsc{fit} task of \textsc{XSPEC} to find a minimum interpolating within the table model. Once this minimum is found, we load the full spectral-timing model into \textsc{pyXSPEC} and we use \textsc{fit} task again to find a more refined minimum. Since this model evaluation is quite slow, calculating error bars for each parameter for each observation using either the \textsc{error} or \textsc{chain} commands in \textsc{XSPEC} is not computationally feasible. Hence, finally, based on this minimum, we use again the table model to calculate the 1-$\sigma$ (68\%) confidence range of each parameter of the model.

\begin{table}
 \caption{Parameters used in the Table Models.}
 \label{tab:table}
 \centering
 \begin{tabular}{c|ccc}
     \hline
     Parameter & Range & Steps & Scale \\
     \hline
     $\nu_{\rm QPO}$~[Hz] & 0.25--6.00 & 17 & linear \\
     $kT_s$~[keV]   & $0.03-2.0$ & 15 & logarithmic \\
     $kT_e$~[keV]   & $5-40$ & 10 & logarithmic \\
     size $(L)$~[km] & $100-25\,000$ & 20 & logarithmic \\
     $\Gamma$ & $1.8-2.8$ & 5 & linear \\
     $\eta$   & $0.001-0.999$ & 21 & linear \\
     \hline
 \end{tabular}
\end{table}

\subsection{Model limitations and caveats}

The model has limitations and caveats that have to be taken into account to avoid over-interpretation of the results inferred from its application, and to understand how the model can be extended and improved, based on the results that will be shown in this paper. In particular, the assumed spherically-symmetric blackbody soft-photon source, is a rough approximation for a spectrum originated in an accretion disc. Furthermore, a spherically symmetric corona with constant optical depth is likely not an accurate representation of the Comptonising region in these sources, which could consist of a non-spherical corona and a relativistic jet. Nonetheless, the quantities inferred by fitting our spectral-timing model to the data of the Type-C QPO can be thought as characteristic values that allow us to constrain physical properties of the system which are not directly accessible through the time-averaged spectra. This way, we can also identify the weaknesses of the model, that will help to pave the road to a more realistic model. 

Future improvements of the model will include a disc-blackbody spectrum for the soft-photon source (Bellavita et al. 2022, in prep.), as a well as a more realistic corona considering gradients of density, temperature and optical depth. For instance, a more complex geometry of the corona could account for the possible the dependence of the rms amplitude and lags upon source inclination \citep{2015MNRAS.447.2059M,Heil15,2017MNRAS.464.2643V}.

The model is based on the solution of the linearised time-dependent Kompaneets equation for Comptonisation, and thus requires that the amplitudes of the oscillations of the temperatures involved ($\delta kT_s$ and $\delta kT_e$) being relatively small to ensure the validity of the perturbative approach. The inferred amplitudes of the oscillating temperatures of the \obsnumber best-fitted observations are shown in detail in Appendix~\ref{app:linearscheme}. We demonstrate that for this particular dataset, $\delta kT_e$ and $\delta kT_s$ remain below 9\% and 5\%, respectively. On the contrary, the amplitude of the external heating rate, $\delta H_{\rm ext}$, can be larger than 100\%, but if a larger emitting area would be considered for the soft-photon source, those values could in principle decrease (for a more detailed discussion, see Appendix.~\ref{app:linearscheme}).

\section{Results}
\label{sec:results}

\begin{figure*}
  \includegraphics[angle=-90,width=0.67\columnwidth]{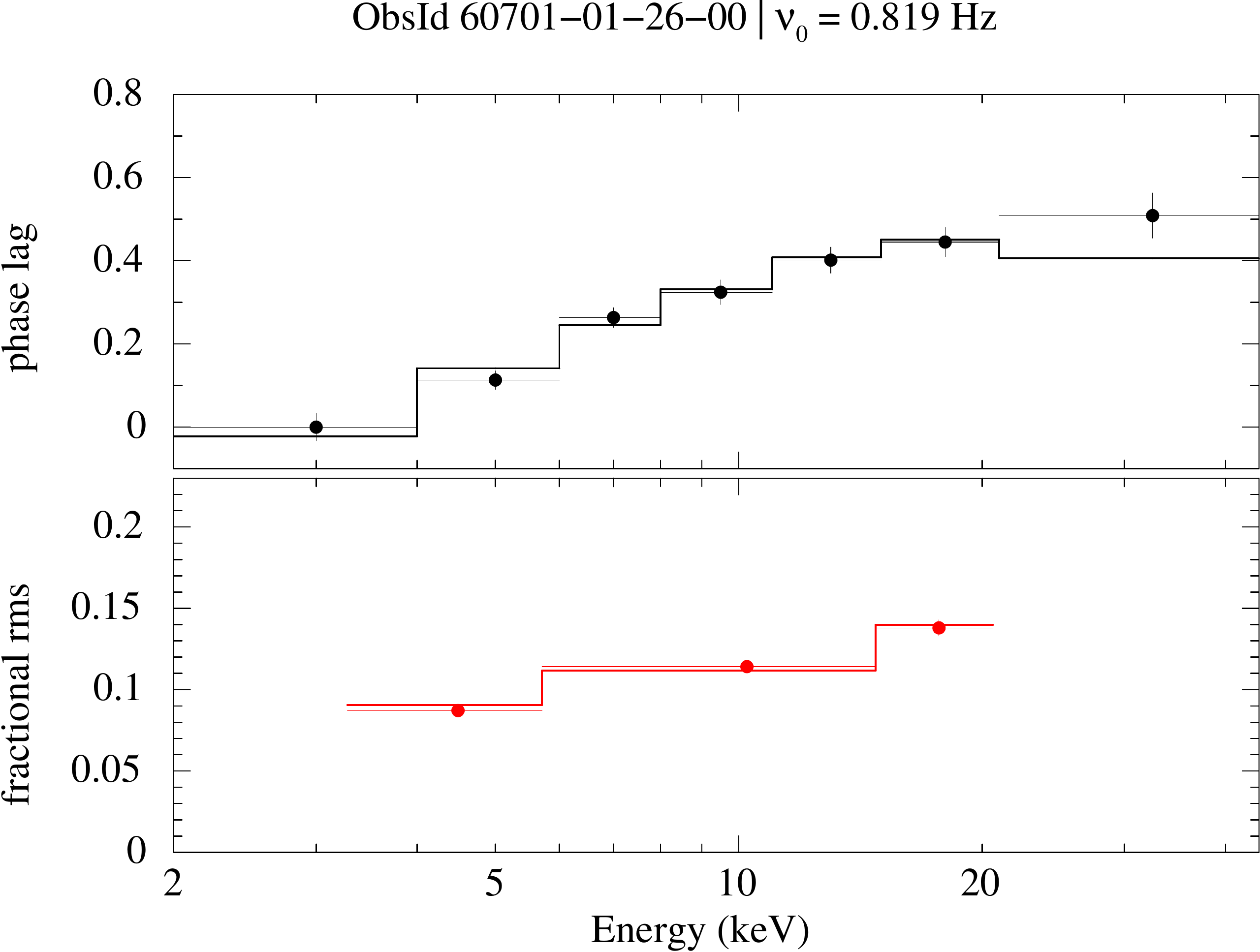}\,\,\,\,
  \includegraphics[angle=-90,width=0.67\columnwidth]{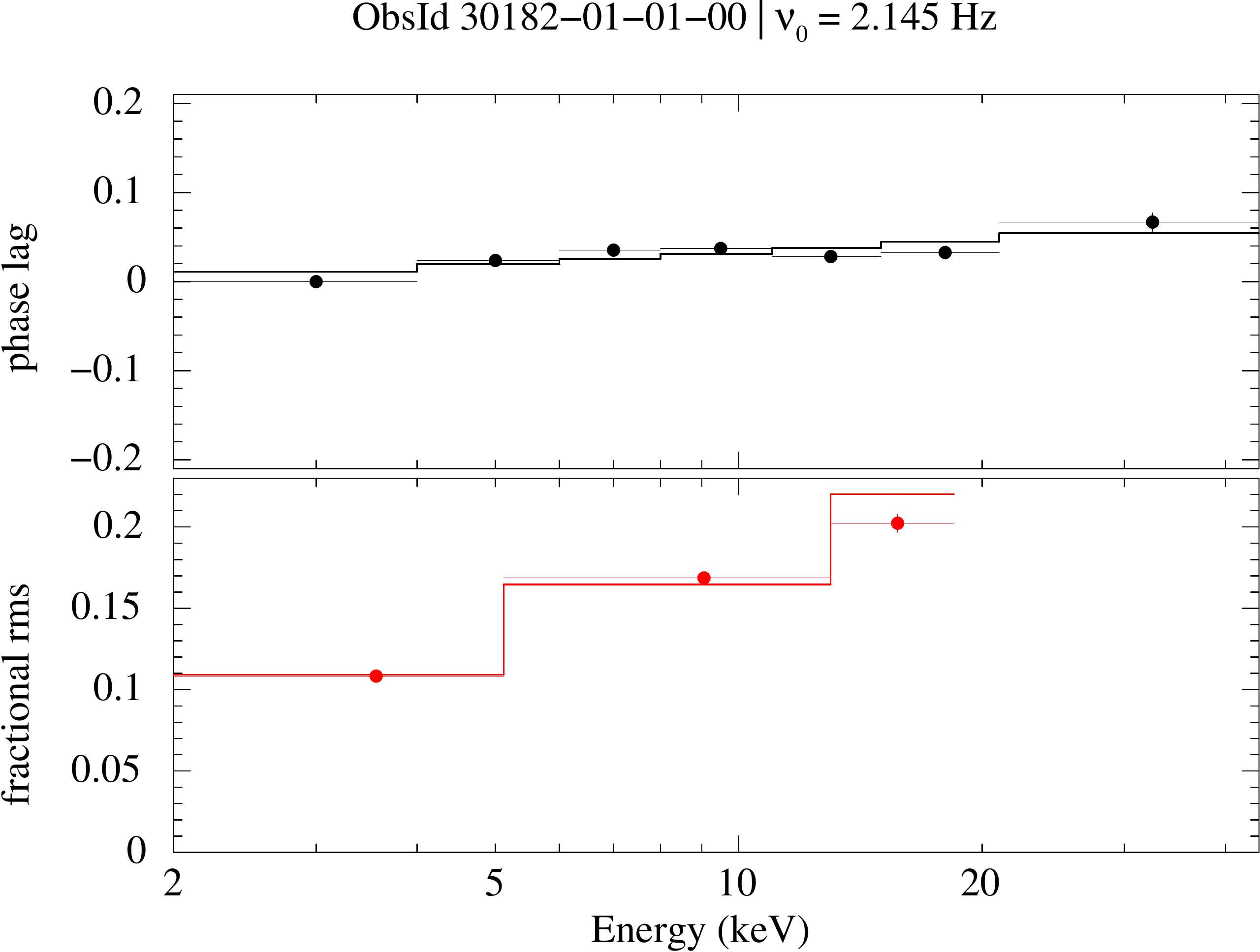}\,\,\,\,
  \includegraphics[angle=-90,width=0.67\columnwidth]{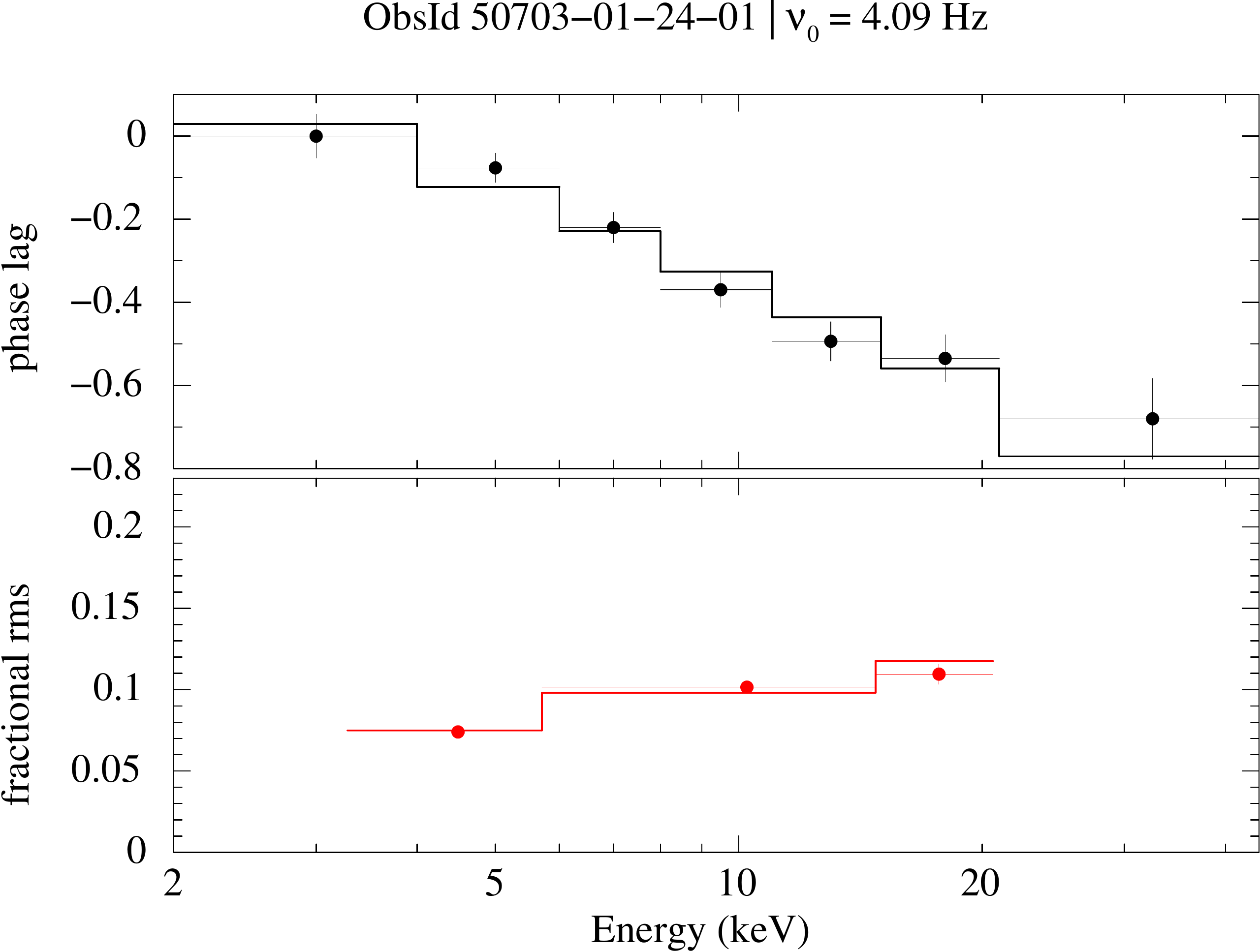}
 \caption{Energy-dependent phase-lag (in radians units, upper panels) and fractional-rms (lower panels) for three \RXTE observations at different QPO frequencies. The left panels correspond to $\nu \approx 0.819$~Hz (below 2~Hz, showing hard lags), central panels to $\nu \approx 2.145$~Hz with zero or flat lags, and the right panels to $\nu \approx 4.090$~Hz (above 2~Hz, showing soft lags).
 The lag- and rms-energy spectra are well-fitted by the spectral-timing Comptonisation model, in the $\sim$2--40~keV and $\sim$2--20~keV energy ranges, respectively.}
 \label{fig:fitting}
\end{figure*}

In Figure~\ref{fig:fitting} we show three examples of the spectral-timing fits. In the left panels we show the results for the QPO in observation 60701-01-26-00, with a $\nu = 0.819$~Hz, which shows hard lags, in the middle panels we show observation 30182-01-01-00, with a QPO at $\nu = 2.145$~Hz, with flat (nearly zero) lags, and in the right panels we show observation 50703-01-24-01 with a QPO at $\nu = 4.09$~Hz, with a soft lag spectrum. In all three cases the top  and bottom panels show, respectively, the energy-dependent phase-lag and fractional rms amplitude spectra. Horizontal error bars correspond to the energy-channel widths and vertical error bars to 1-$\sigma$ (68\%) uncertainties in the underlying data. Solid lines in these plots represent the best-fitting spectral-timing model. We notice that, while the shape of the phase-lag spectra changes significantly between the three panels, showing either positive, flat or negative slopes, the rms-amplitude energy spectra are more or less similar, and  always show an increasing rms with energy. The model can fit both rms- and lag-energy spectra simultaneously, with the differences in the slope of the lag-energy spectra being driven mainly by the feedback fraction, as we explain in more detail below. The residuals of the \obsnumber best-fitting models are presented in detail in Appendix~\ref{app:goodness}.

\begin{figure*}
 \includegraphics[width=\columnwidth]{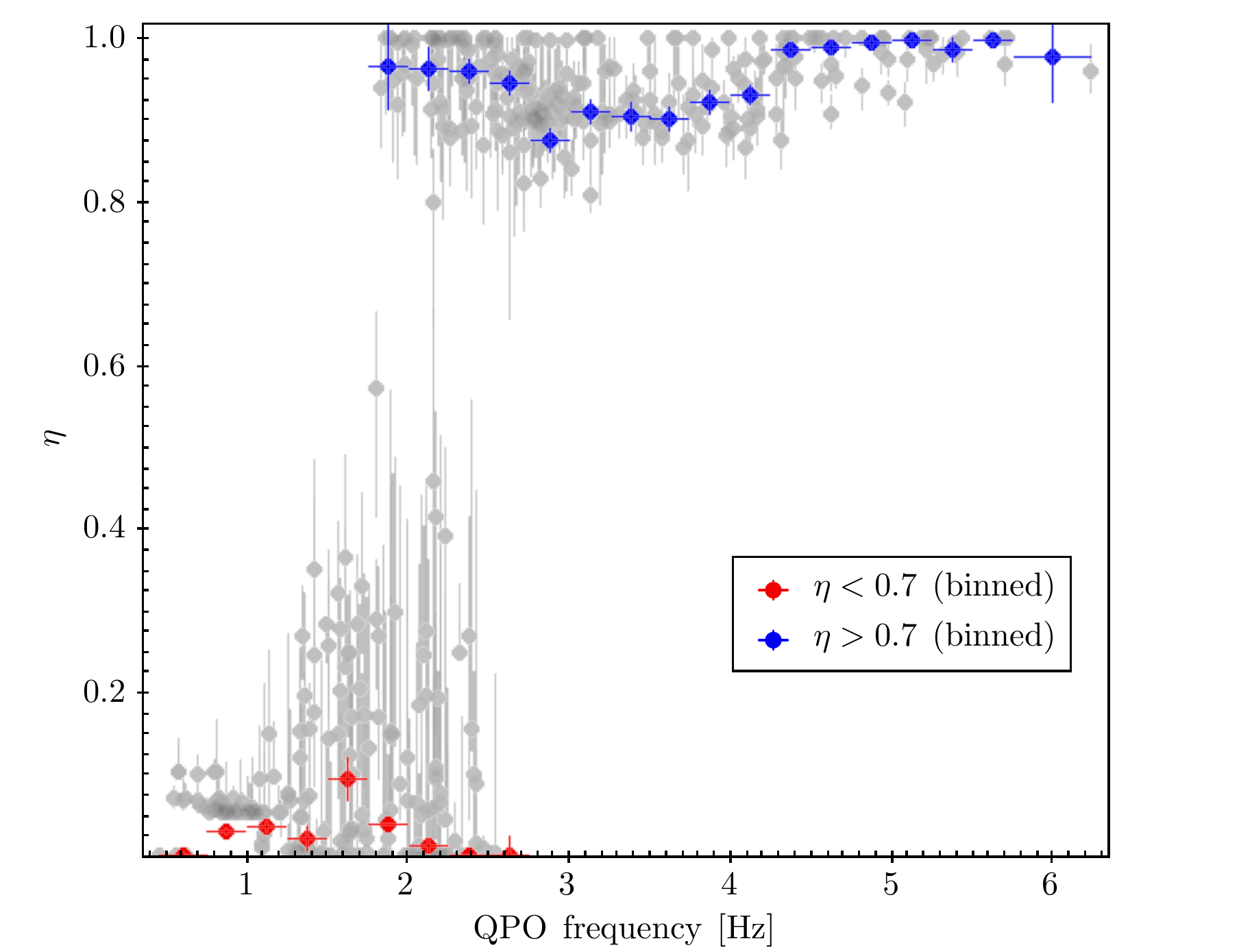}\,\,\,\,
 \includegraphics[width=\columnwidth]{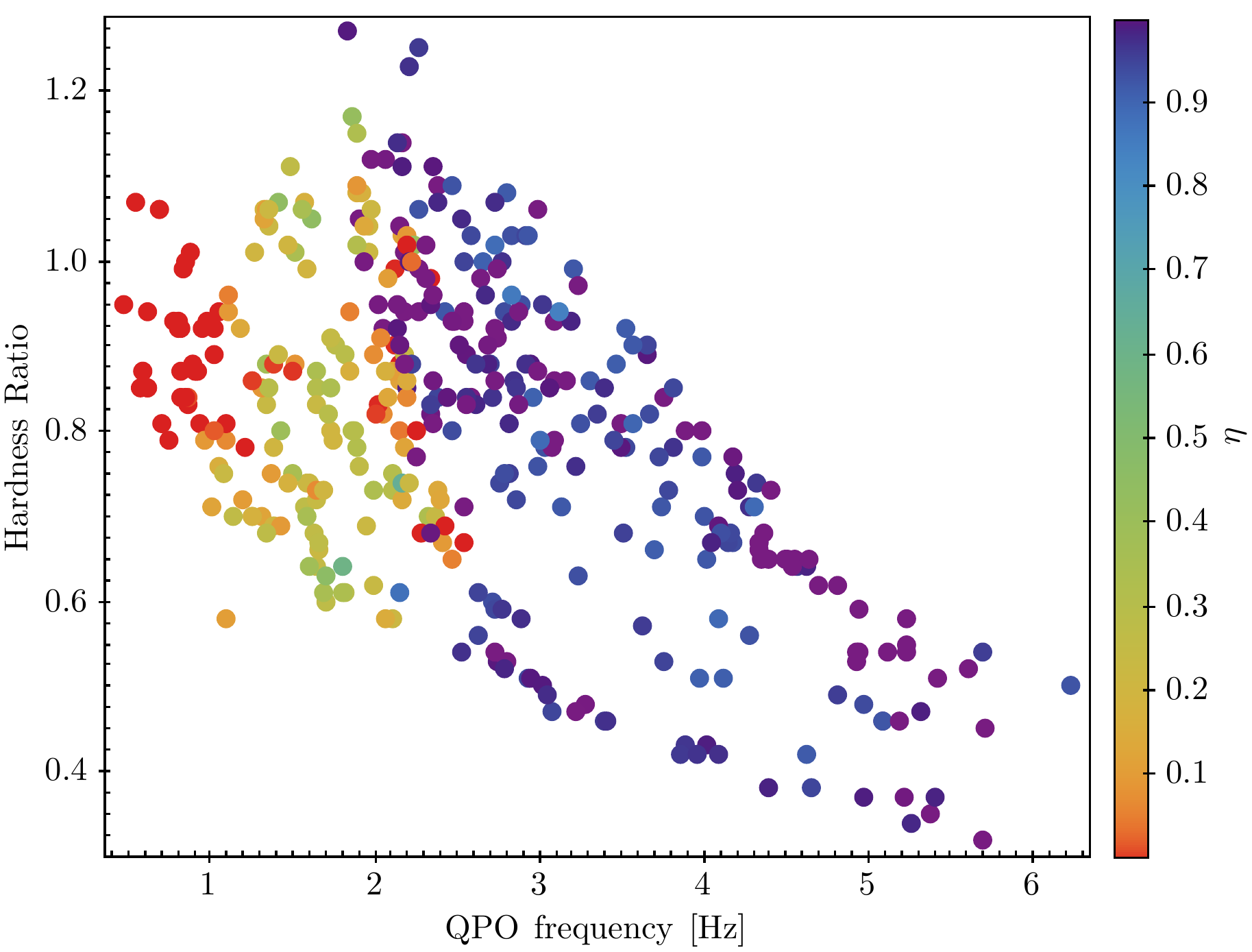}
 \caption{{\bf Left panel:} feedback fraction, $\eta$, and its 1-$\sigma$ uncertainty, as a function of QPO frequency. Each grey point corresponds to one of the \obsnumber observations considered in this work. Data is also weighted-averaged in frequency bins every 0.25~Hz considering separately those data points with $\eta > 0.7$ (blue colour) and $\eta < 0.7$ (red colour). {\bf Right panel:} $QPO-HR$ diagram of the feedback fraction.}
 \label{fig:eta}
\end{figure*}

\subsection{Dependence on feedback fraction}
\label{sec:feedback}

In the left panel of Figure~\ref{fig:eta} we show the best-fitting values to the feedback fraction parameter, $\eta$, and their corresponding 1-$\sigma$ error bars as a function of the QPO frequency for the full data set (grey colours). Globally, the feedback fraction clearly shows two different cluster families: {\em a}) for $\nu \la 1.8$~Hz, the feedback is low ($\eta < 0.2-0.3$); {\em b}) for $\nu \ga 2.5$~Hz, the feedback is high ($\eta > 0.7-0.8$), and shows an increasing trend as $\nu_{\rm QPO}$ increases; {\em c}) for $\nu \sim 1.8-2.5$~Hz there is a transition from one regime to the other, and both clusters coexist. By comparing with the behaviour of the lags (see left panel of Fig.~\ref{fig:lags_and_rms}), we find that regime {\em (a)} corresponds to the QPO frequencies where the lags are hard, and {\em (b)} corresponds to the regime where the lags are soft. Meanwhile, flat or {\em zero} lags are found in the intermediate regime {\em (c)}, where the feedback can be either low or high. Given the large amount of observations fitted, the spread of the points and the relatively large error bars, we calculated the weighted-average of this parameter according to the QPO frequency into 4 bins per Hz. We produce two sets of binned data points: i) points with $\eta <0.7$ (red colour, and consistent with regime {\em (a)}, ii) points with $\eta>0.7$ (blue colour, consistent with regime {\em (b)}.
The binned points make evident that, notwithstanding the regime {\em (c)} ($\nu_0 \sim 1.8-2.5$~Hz) where the uncertainties are the largest, and both clusters coexist, regimes {\em (a)} and {\em (b)} have a persistent trend, where $\eta$ increases as the QPO frequency increases. 
A first indication of this trend was presented in \citet{2021MNRAS.503.5522K} using nine observations. Based on the full set of \obsnumber observations available in the \RXTE archive, we recover that this behaviour persists on average, but we also find that regimes {\em (a)} and {\em (b)} coexist in the 2--3~Hz intermediate frequency range, a result that was hard to achieve with the sub-sample of nine observations analysed there.

In the right panel of Figure~\ref{fig:eta} we use coloured points to represent the values of $\eta$ in a $QPO-HR$ diagram. Here it is apparent that the transition from low to high $\eta$ fractions at $\sim$2~Hz is independent of the $HR$ state of the source, and occurs at the QPO frequency at which the lags turn from hard to soft (see left panel of Fig~\ref{fig:lags_and_rms}). Despite the uncertainties principally present around 2--3~Hz, where the error bars in $\eta$ are the largest, a monotonic relation between $\eta$ and $\nu_{\rm QPO}$ can be readily seen in this plot. In particular, the largest ($\eta \approx 1$) values tend to appear mainly at the highest QPO frequencies, at the bottom-right corner of the plot.

\begin{figure*}
 \includegraphics[width=\columnwidth]{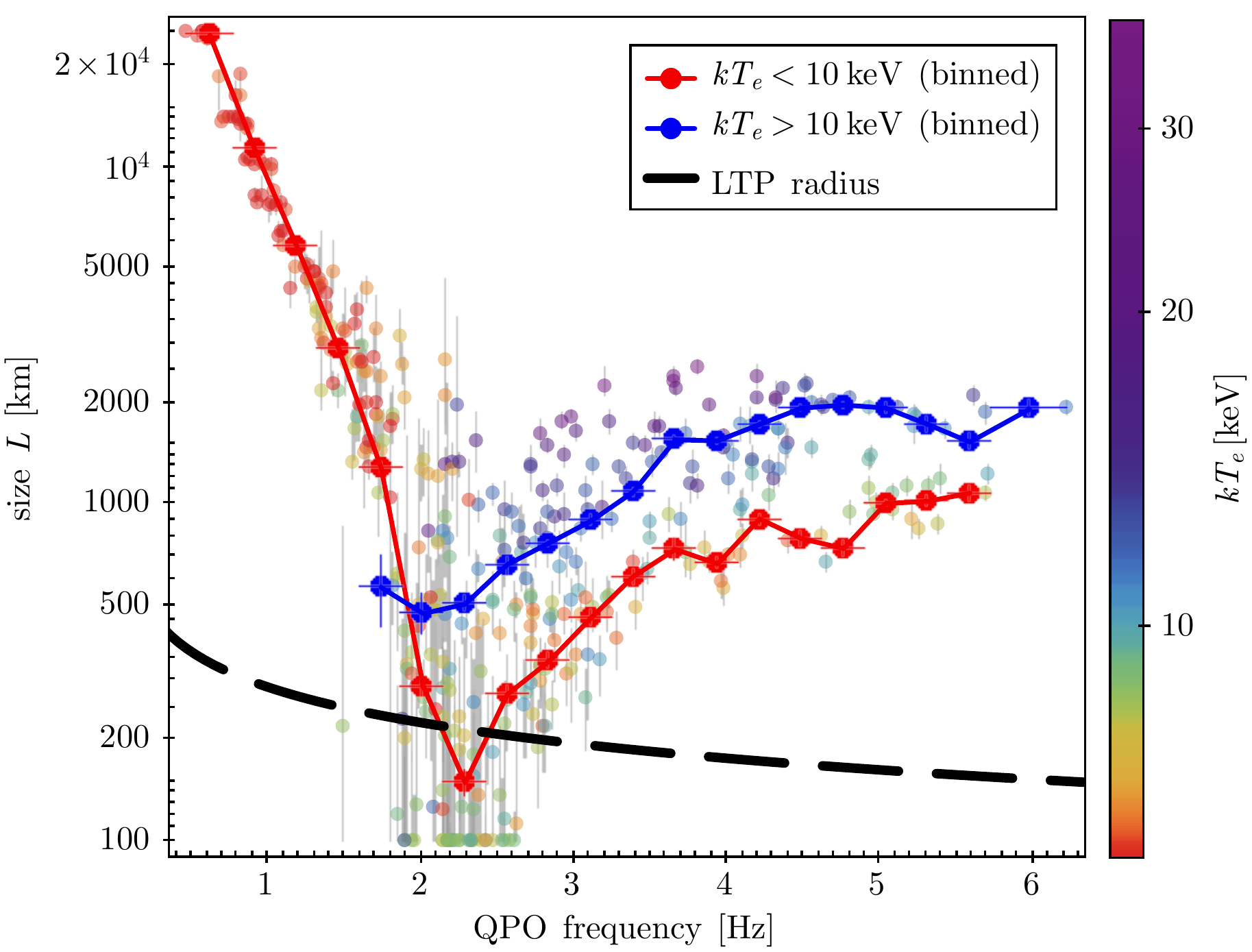}\,\,\,\,
 \includegraphics[width=\columnwidth]{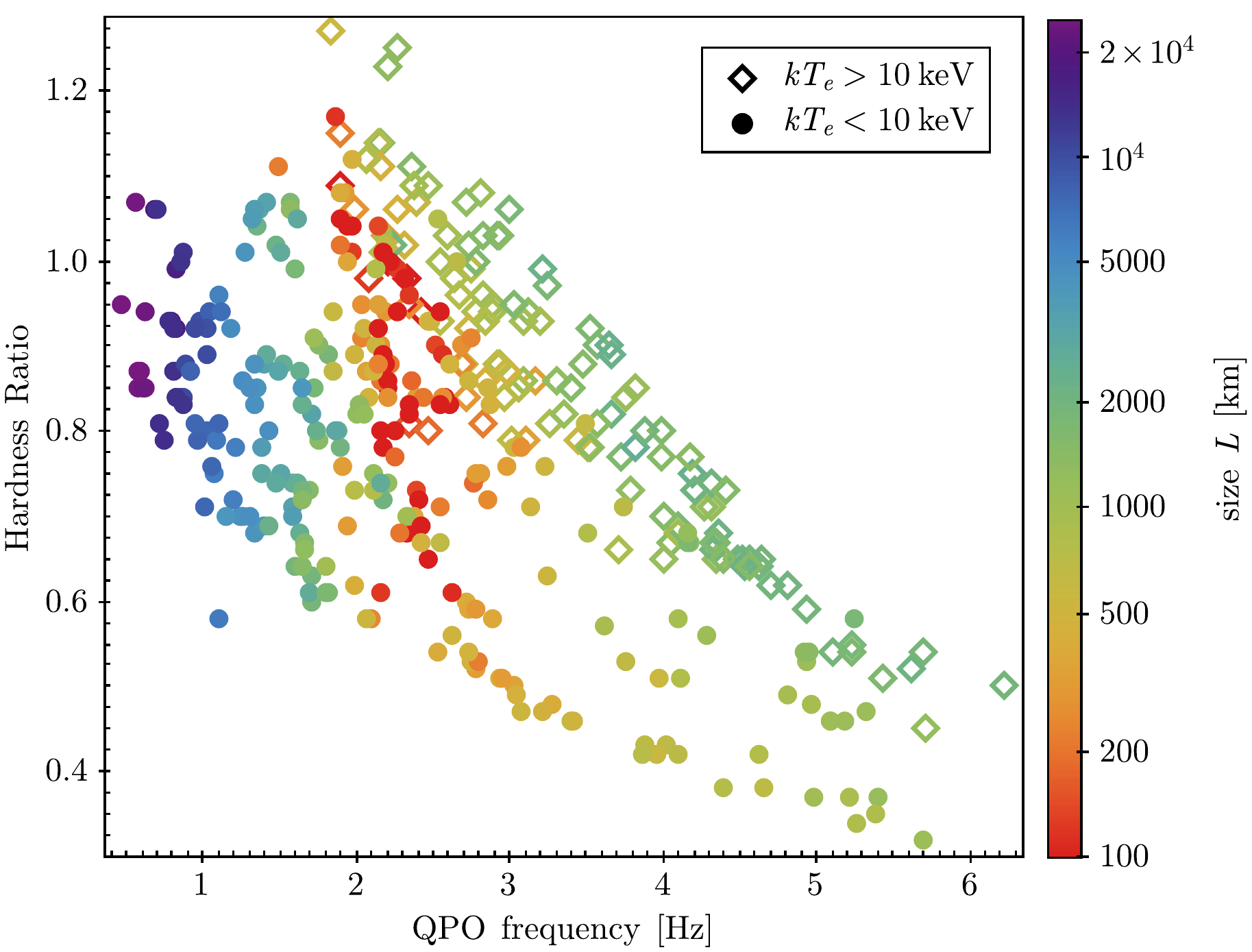}
 \caption{{\bf Left panel:} corona size, $L$, and its 1-$\sigma$ uncertainty, as a function of QPO frequency. Small points with grey error bars correspond to the \obsnumber observations considered in this work. Those points are coloured based on the temperature of the corona, $kT_e$ (see colourbar). 
 Blue and red colours correspond to weighted-averaged observations with $kT_e > 10$~keV and $kT_e < 10$~keV, respectively. Dashed-black curve represents the radius associated to the LTP frequency for a $\sim$12~M$\odot$ BH, as in \GRS. 
 {\bf Right panel:} $QPO-HR$ diagram of the corona size (indicated with colours according to the colour bar). Solid points (empty diamonds) are used for observations with a cool (hot) corona of $kT_e < 10$~keV ($kT_e > 10$~keV). Our multi-dimensional data set allows us to unravel the changes in the physical properties of the source that lead to this separation.}
 \label{fig:size}
\end{figure*}

\subsection{The corona size behaviour}
\label{sec:size}

In the left panel of Figure~\ref{fig:size} we show the best-fitting values to the corona size, $L$ in km, and their corresponding 1-$\sigma$ error bars, as a function of QPO frequency, for each of the \obsnumber observations considered. To give the reader an idea of the dimensions associated to $L$ in physical units, we also plot a black dashed line on top corresponding to the LTP radius for each QPO frequency (considering a $\sim$12~M$_\odot$ BH, like \GRS), which is usually associated to the inner edge of the accretion disc, $r_t$ \citep{2009MNRAS.397L.101I}. In the 0.5--6.5~Hz range, the LTP radius spans from 350~km ($\sim$20~$R_g$) down to 150~km ($\sim$8~$R_g$). Focusing on the points corresponding to the best-fitting sizes, at low frequencies ($\nu \la 2$~Hz) $L$ rapidly decreases as the QPO frequency increases, going from $\ga$10\,000~km when $\nu \la 0.5$~Hz, to $\la$1000~km for $\nu \la 2$~Hz. In the 2--3~Hz range the corona reaches minima sizes becoming compatible at the 1-$\sigma$ level with the minimum size explored in the table models (100~km). From this point on wards, the pattern is reversed, at high frequencies ($\nu \ga 3$~Hz), $L$ has more or less constant values spanning from $\sim$500 to 2000~km with a tendency to increase as the QPO frequency increases. Moreover, by colouring these points based on the electron temperature in the corona, taken from the time-averaged spectra, a hint for two families of $L$ values can be seen in the data in this latter frequency regime.

To test this possibility, we calculate the weighted-average of the data in frequency bins of 0.25~Hz, considering two different subsets according to their $kT_e$ values, arbitrarily chosen either below 10~keV (red colour) or above (blue colour). Two clear monotonic, but separate, trends are apparent: while at $\nu \ga 2$ Hz the corona size increases as QPO frequency increases, consistent with the results of \cite{2021MNRAS.503.5522K}, we find that the size is systematically larger when the corona is hot ($>$10~keV) than when it is cool ($<$10~keV). This becomes more evident on the right panel of Figure~\ref{fig:size}. Here we colour the points of the $QPO-HR$ diagram according to the best-fitting value of the corona size. An almost horizontal gradient is apparent, showing the dependence of $L$ with the QPO frequency described above. However, another more subtle relation arises in this plot. When $\nu \ga 3$~Hz, $L$ is systematically larger on the top-right band of points (greenish empty diamonds, $kT_e > 10$~keV) than in the bottom band points (brown-yellow solid circles, $kT_e < 10$~keV). Those points have significantly different $kT_e$ values (see left panel of Fig.~\ref{fig:kTe_and_Gamma}), and thus, the two families of $L$ values on that regime can be explained by different thermodynamic conditions of the corona at the same QPO frequency: the higher $kT_e$, the larger $L$.

\begin{figure*}
\includegraphics[width=\columnwidth]{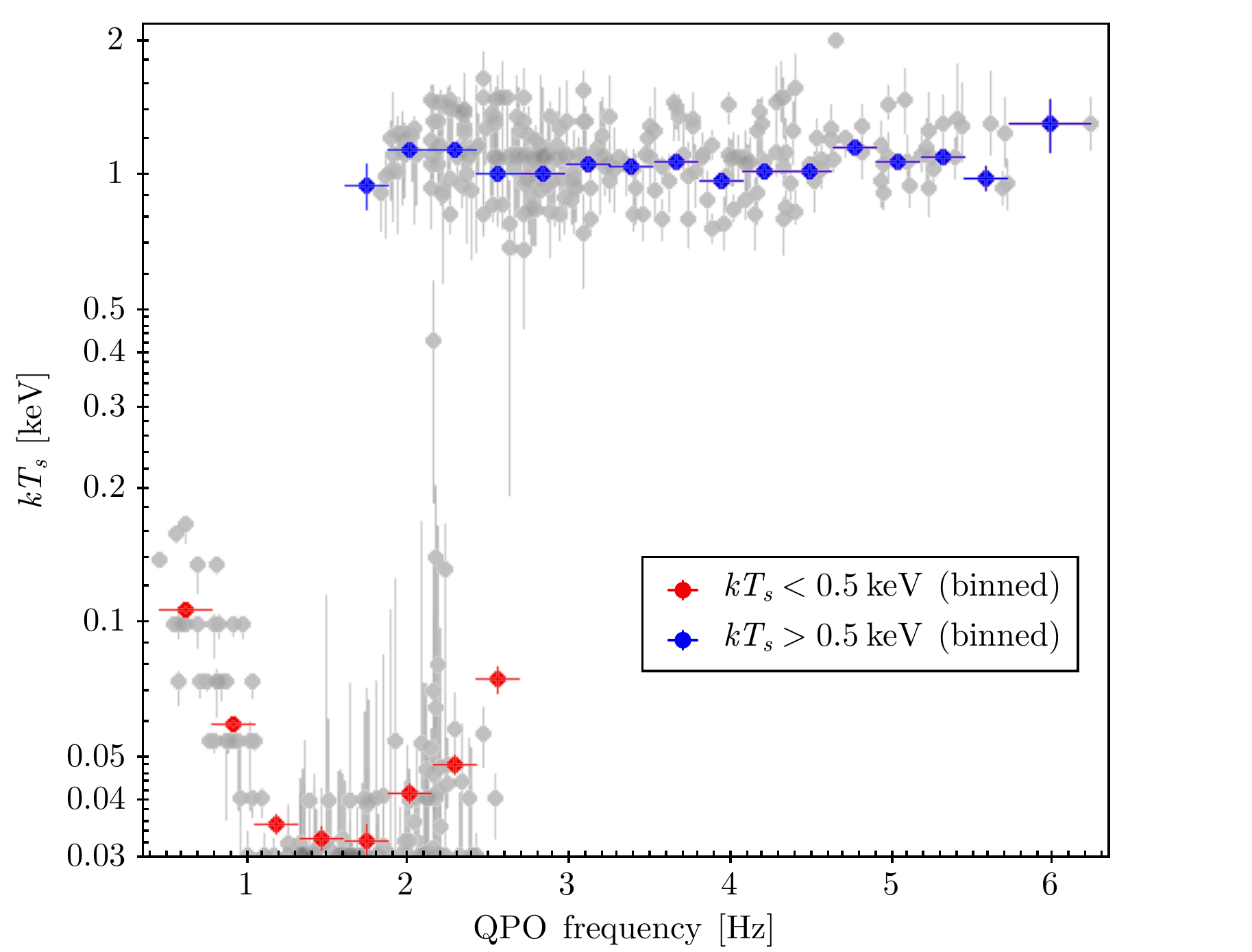}\,\,\,\,
 \includegraphics[width=\columnwidth]{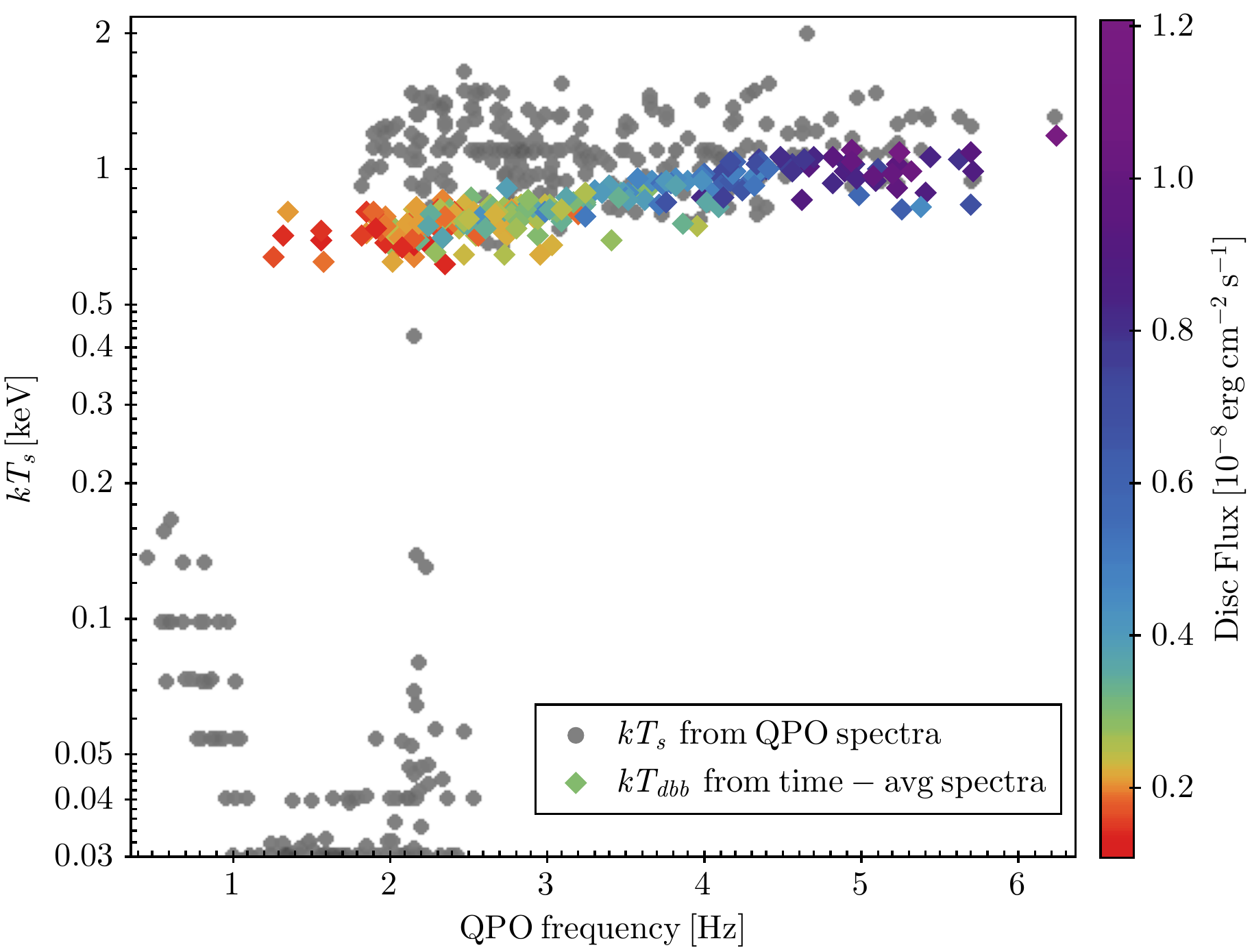}
 \caption{{\bf Left panel:} temperature of the soft-photon source, $kT_s$, and its 1-$\sigma$ uncertainty, as a function of QPO frequency (grey points and error bars). Data is also weighted-averaged in frequency bins every 0.25~Hz. Blue and red colours correspond to weighted-averaged observations with $kT_s > 0.5$~keV and $kT_s < 0.5$~keV, respectively. 
 {\bf Right panel:} Comparison between the soft-photon source temperature estimations using the different data available. Grey points (as in the left panel) correspond to the temperature of the black-body like source fitted using the spectral-timing data of the QPO ($kT_s$), while coloured diamonds are used for the temperature of the disc blackbody fitted using the time-averaged spectra. Those points are coloured according to the flux of the disc (see colour bar). The missing coloured points (for $\nu \la 2$~Hz) correspond to observations where the disc blackbody could not be fitted to the spectra.}
 \label{fig:kTs}
\end{figure*}

\subsection{The soft-photon source temperature}
\label{sec:kTs}

In the left panel of Fig.~\ref{fig:kTs} we show the best-fitting values obtained for the soft-photon source temperature ($kT_s$, grey points) and their 1-$\sigma$ uncertainties. The values of $kT_s$ separate into two different regimes, according to the QPO frequency. For $\nu_0 \ga 2.5$~Hz we find relatively large $kT_s$ values, $0.7<kT_s<2$~keV, whereas for $\nu_0 \la 1.8$~Hz, we obtain low temperatures, $kT_s < 0.2$~keV, even compatible with the minimum values explored with our table models ($kT_s = 0.03$~keV). As in the case of the feedback fraction in Sec.~\ref{sec:feedback}, in the 1.8--2.5~Hz frequency range $kT_s$ can be either in the low- or high-temperature regime, or even with values in between. Following what we did for the feedback fraction, we also calculate the weighted-average of the $kT_s$ values in frequency bins of 0.25~Hz. For this we consider observations with $kT_s$ values either higher (blue points) or lower (red points) than 0.5~keV.
For frequencies $\nu_0 \ga 2.5$~Hz, the temperature tends to increase with QPO frequency (notice that the y-axis in the plot is in logarithmic scale), which is expected if this temperature traces the temperature of the innermost regions of the accretion disc, and the QPO frequency itself corresponds to the LTP frequency at the inner-edge of the accretion disc.
This trend breaks below $\sim$1.8~Hz, where a transition to a low-temperature soft-photon source occurs. This transition becomes more evident when the temperature of the disc black-body component, $kT_{\rm dbb}$, fitted to the time-averaged spectra is also considered \citep{2022mariano}. 

In the right panel of Fig.~\ref{fig:kTs}, we show the comparison between the two soft-photon source temperatures fitted to each observation using the two different data available. 
Grey points (as in the left panel) correspond to the temperature of the blackbody-like source fitted using the spectral-timing data of the QPO ($kT_s$), while coloured diamonds are used for the temperature of the disc blackbody fitted using the time-averaged spectra ($kT_{\rm dbb}$). 
The latter set of points is coloured according to the flux of the disc-blackbody component (see colour bar). Below $\sim$2~Hz, only 6 observations have a disc-blackbody temperature measurement. The missing coloured points correspond to observations where only an upper limit on the disc blackbody component could be given from the fits to the time-averaged spectra, as this component was not significantly present in the fits. In those observations, the best-fitting model is simply \textsc{tbabs*(nthcomp+gaussian)}, in \textsc{XSPEC} notation \citep[we refer the reader to][for more details]{2022mariano}.

We remind here that in order to fit the spectral-timing data of the QPO we have fixed both $kT_e$ and $\Gamma$ to the values fitted to the Comptonisation component in the time-averaged spectra. Since the best-fitting soft-photon temperatures found in this work are compatible with those found by fitting the time-averaged data using the \textsc{nthcomp} model, which coincides with the steady-state solution of our variable-Comptonisation model (see Sec.~\ref{sec:model}), this means that the predicted time-averaged spectra associated to our Comptonisation model are also consistent with the observed ones.

\section{Discussion}
\label{sec:discussion} 

We have successfully applied our spectral-timing model for Comptonisation to the Type-C QPO data of \GRS available in the full \RXTE archive, consisting of \obsnumber observations performed with the PCA camera. By combining the energy-dependent phase-lag spectra of the QPO presented in \cite{2020MNRAS.494.1375Z}, with our own measurements of the fractional-rms spectra (presented in Sec.~\ref{sec:rmslags}), together with information gathered from the time-averaged spectra \citep{2022mariano}, we are able to constrain physical and geometrical properties of the corona, like the feedback fraction, $\eta$, and its characteristic size, $L$, which are not directly accessible through the frequently used time-averaged spectra. Based on this data set, we analyse the dependence of the corona properties on both the QPO frequency and the spectral state of the source, given by the hardness ratio, as well as their mutual interconnections. We identify solid trends in the evolution of the feedback fraction, corona size, and temperature of the Comptonised soft-photon source, which persist during the full time span by the \RXTE mission in space, of about 15~years. This allow us to construct a global picture of the disc-corona interaction and their links with the jet-launching mechanism that we will discuss hereafter.

The outstanding dependence of the slope of the lag-energy spectra with QPO frequency \citep{2020MNRAS.494.1375Z} can be interpreted in the context of our spectral-timing Comptonisation model by two families or clusters of solutions. One having a large feedback fraction ($\eta > 0.8$) and high soft-photon source temperatures ($kT_s \sim 0.8-1.5$~keV), when $\nu_0 > 2.5$~Hz, and negative or soft lags; and the other having low feedback fraction ($\eta < 0.5$) and low soft-photon source temperatures  ($kT_s < 0.2$~keV), when $\nu_0 < 1.8$~Hz and positive or hard lags. Each of these clusters show a smooth trend, with increasing $\eta$ and $kT_s$ as $\nu_0$ increases. Furthermore, in the $\sim$1.8--2.5~Hz range, where the lags are approximately flat or zero, we find that both clusters of solutions coexist, and a sharp transition between both states is recovered. When the dependence on $HR$ is explored, no clear trend is found for any of these two properties.

Remarkably, the relation of $kT_s$ with QPO frequency is also independently supported by the time-averaged spectra. As shown in the right panel of Fig.~\ref{fig:kTs}, while observations with high $kT_s$ also show a significant \textsc{diskbb} component in the time-averaged spectrum, with systematically increasing $kT_{\rm dbb}$ and disc flux, with increasing $\nu_0$, this is not the case when $\nu_0 \lesssim 2$~Hz. In this latter regime, only a few (6) observations show a significantly detected \textsc{diskbb} component, while in the vast majority the continuum is fitted solely with an \textsc{nthcomp} component \citep{2022mariano}. Moreover, the connection between the feedback fraction and soft-photon source temperature is expected in our model, given that if a large fraction of photons impinge back onto the soft source, the source will be heat-up and thus experience a higher $kT_s$ temperature, and {\em vice versa}. 

The corona size behaviour is more complex and rich. For QPO frequencies $\nu_0 \lesssim 2$~Hz, the size of the corona consistently shrinks from $L \gtrsim 10^4$~km ($\sim$500~$R_g$) to $L \sim 200-500$~km ($\sim$10--25~$R_g$) as $\nu_0$ increases from $\sim$0.5 to 2~Hz. From this frequency onwards, the corona consistently expands again to sizes of $\sim$1000--2000~km (50--100~$R_g$) when the QPO frequency reaches its maximum observed values $\sim$6~Hz. In this regime, the $L-\nu_0$ evolution shows a much broader relation, with different corona sizes coexisting at the same QPO frequency. When the corona temperature, $kT_e$, is taken into account, the breath of this relation can be explained as a connection between the thermodynamic properties of the corona and its geometry. At the same QPO frequency, systematically larger corona sizes are found for higher $kT_e$ values (see Fig.~\ref{fig:size}). 

While in the variable-Comptonisation model the feedback has the largest impact upon the sign of the slope of the lag spectrum (either positive or hard, for very low $\eta$, or negative or soft for high $\eta$), the magnitude of the slope of the lags is driven by the corona size: i.e., for the same constant optical depth, in a larger corona photons are up-scattered more times than in a smaller one. For this reason, for a very small corona ($L$ of a few gravitational radii) flat (or zero) lags are expected. Thus, given the strong connection between the QPO frequency and the magnitude of the phase-lags in \GRS \citep{2020MNRAS.494.1375Z}, the appearance of trends like those found for the dependence of the corona size with the QPO frequency are naturally expected. Moreover, the dependence of these trends on the temperature of the corona $kT_e$, can also be understood based on the mathematical relationship expressed by Eq.~(\ref{eq:gamma}) between $kT_e$, the spectral index, $\Gamma$, and the optical depth, $\tau$.

Assuming that the QPO frequency traces the LTP radius of the inner edge, $r_t$, of the accretion disc \citep{2009MNRAS.397L.101I}, the large feedback experienced by the source when $\nu_0 > 2.5$~Hz, can be explained by a corona that fully enshrouds the inner regions of the accretion disk with sizes $L >> r_i$. Remarkably, for frequencies $\nu \sim 2$~Hz, the corona shrinks to minima sizes $L \sim r_t$, leading to flat lags that can be fitted by either large or small $\eta$ values, and high or low $kT_s$ temperatures, respectively. We interpret this as the crossing between the corona size and the inner radius of the disc, which turns from a strong feedback process to a very inefficient one, when the corona becomes smaller than the inner edge of the disc. As the QPO frequency decreases even further ($\nu_0 < 1.8$~Hz) two important changes occur: the corona becomes cooler $kT_e \sim 5-8$~keV and the $L - \nu_0$ relation reverses, but keeping the feedback fraction low. This could be explained by a change in the geometry, in which the corona becomes extended perpendicular to the plane of the accretion disc, in a jet-like manner, avoiding the formation of a strong feedback process, and thus leading to positive lags.

\begin{figure}
 \includegraphics[width=\columnwidth]{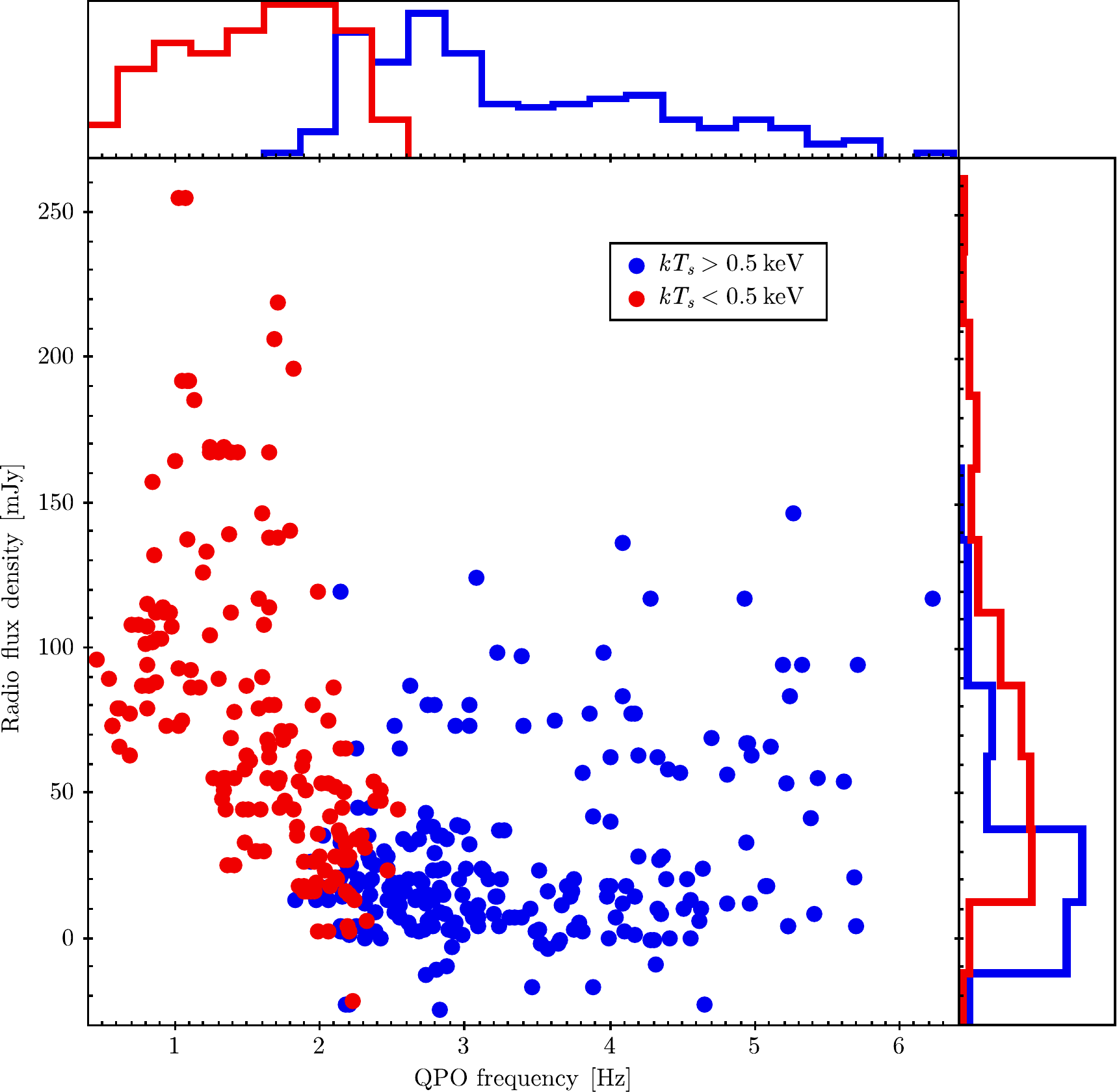}\\
 \caption{{\bf Main panel:} Radio flux density (Ryle measurements at 15~GHz) as a function of the QPO frequency at each corresponding $\RXTE$ observation. Red (blue) coloured points correspond to temperatures of the soft-photon source, $kT_s < 0.5$~keV ($kT_s > 0.5$~keV), fitted to the spectral-timing data of the QPO. Average errors are $\pm$(1--5)~mJy and $\pm$0.05~Hz, in radio measurements and QPO frequency, respectively. {\bf Top panel:} normalised marginal distributions of QPO frequencies for each subset. Observations with $kT_s < 0.5$~keV ($kT_s > 0.5$~keV) accommodate below (above) $\sim$2~Hz. {\bf Right panel:} normalised marginal distributions of radio intensities for each subset. Observations with $kT_s < 0.5$~keV have systematically higher radio flux density values than those with $kT_s > 0.5$~keV. }
 \label{fig:radio}
\end{figure}

Based on the full data set of \RXTE X-ray observations containing a significant Type-C QPO in their PDS, we collected and cross-matched pointed radio observations performed by the Ryle telescope at 15~GHz \citep{1997MNRAS.292..925P} within 2~days of the corresponding X-ray observations (leading to a total sample of \obsnumber data points). The radio measurements span from very low values, or non-detections (with flux-density levels $\la$5--10~mJy) to very significant detections in the range of 10--250~mJy (with average uncertainties of $\pm$1--5~mJy). In the main panel of Fig.~\ref{fig:radio} we show the radio flux density corresponding to each of the \obsnumber observations as a function of QPO frequency. The points are coloured according to the temperature of the soft-photon source fitted to with the spectral-timing model, $kT_s$. Red (blue) coloured points correspond to $kT_s < 0.5$~keV ($kT_s > 0.5$~keV). On the top and right panels we show the normalised marginal distributions of QPO frequencies and radio flux density values for each subset, respectively. 

Figure~\ref{fig:radio} shows an anti-correlation between radio flux density and the QPO frequency. In the same Figure, a correlation between the QPO frequency and the temperature of the soft-photon source is also evident. For each X-ray observation where the temperature of the soft-photon source is low ($kT_s < 0.5$~keV, red colour), the source is significantly detected in radio with flux densities in the 65$^{+54}_{-38}$~mJy range (1-$\sigma$). Meanwhile, in observations for which the temperature is high ($kT_s > 0.5$~keV, blue colour), the radio fluxes are significantly lower, spanning in the 18$^{+44}_{-15}$~mJy range (1-$\sigma$), with a significant fraction of the observations being consistent with non-detections (see right panel). A Kolmogorov-Smirnov test between the two samples of radio measurements throws a probability of $\sim$10$^{-16}$, meaning that the null hypothesis that the two samples come from the same underlying distribution can be rejected with >8-$\sigma$. Such correlation was firstly suggested by \cite{2001ApJ...558..276T} and later by \cite{2013MNRAS.434...59Y}, using a much smaller data set of \RXTE observations, that we have here extended to the full \RXTE archive. In addition, the separation of these two populations in terms of QPO frequency is very clear: points with $kT_s < 0.5$~keV (red colour) and $kT_s > 0.5$~keV (blue colour) accommodate either below or above $\approx$2~Hz, respectively (see top panel).

We finally argue that a possible change in the actual source of soft photons when $\nu_0 < 2$~Hz can not be discarded. Given that the disc-blackbody component becomes insignificant in the time-average spectra; the low $kT_s$ temperatures inferred from our spectral-timing Comptonisation model; the very-long positive lags, that lead to large inferred corona sizes; and the consistent appearance of significant radio detections associated to synchrotron emission in the compact jet, this could mean that the dominating source of soft photons for Comptonisation in X-rays could become the synchrotron photons of the radio jet, in the so-called synchrotron self-Compton radiative process \citep{2009MNRAS.392..570M,2009ApJ...690L..97P}. In this case, the corona, i.e. the Comptonising region, would become the base of the jet, as proposed by \cite{1999MNRAS.304..865F} and discussed in more detail in \cite{2005ApJ...635.1203M}. The possibility that the Comptonising cloud is effectively ejected as a jet giving place to a radio flare has been suggested by \cite{2003ApJ...597.1023V}, \cite{2003ApJ...595.1032R}, and \cite{2004MNRAS.355.1105F}. Results from 15~years of simultaneous X-ray observations with \RXTE and radio-monitoring with the Ryle telescope provide very strong evidence of a direct coupling between the X-ray corona and the radio jet \citep{2022mariano} in \GRS.

\section*{Acknowledgements}

We thank the Referee for insightful comments that helped us to improve this paper. We are grateful to M. Taylor for providing a tailored colour bar scaling for \textsc{topcat} \citep{2005ASPC..347...29T}. This work is part of the research programme Athena with project number 184.034.002, which is (partly) financed by the Dutch Research Council (NWO). FG acknowledges support by PIP 0113 (CONICET). FG is a CONICET researcher. This work received financial support from PICT-2017-2865 (ANPCyT). LZ acknowledges support from the Royal Society Newton Funds. YZ acknowledges support from China Scholarship Council (CSC 201906100030). TMB acknowledges financial contribution from the agreement ASI-INAF n.2017-14-H.0, PRIN-INAF 2019 N.15, and thanks the Team Meeting at the International Space Science Institute (Bern) for fruitful discussions. DA acknowledges support from the Royal Society.

\section*{Data Availability}

This research has made use of data obtained from the High Energy Astrophysics Science Archive Research Center (HEASARC), provided by NASA’s Goddard Space Flight Center. The radio data used in this study are available at \url{https://www.astro.rug.nl/~mariano/GRS_1915+105_Ryle_data_1995-2006.txt}.



\bibliographystyle{mnras}
\bibliography{2022_GRS1915_FG} 

\appendix

\section{External heating rate: validity of the linearised Kompaneets equation}
\label{app:linearscheme} 

\begin{figure}
 \centering
 \includegraphics[width=0.9\columnwidth]{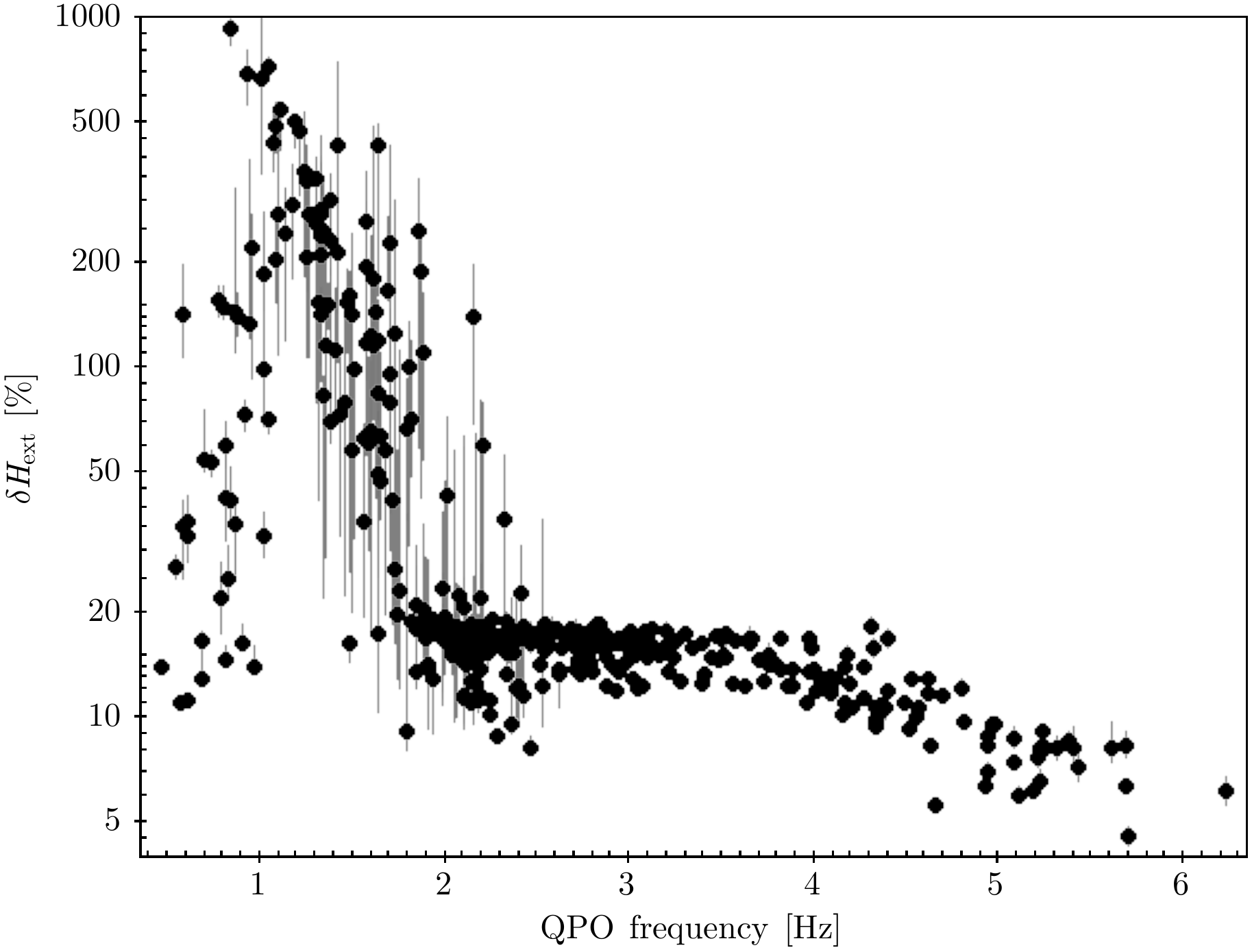}
 \caption{Amplitude of the variation of the external heating rate, $\delta H_{\rm ext}$, as a function of QPO frequency for the \obsnumber observations.}
 \label{fig:dHext}
\end{figure}

In Fig.~\ref{fig:dHext} we show the best-fitting values, and corresponding 1-$\sigma$ error bars, obtained for the amplitude of the external heating rate source, $\delta H_{\rm ext}$, as a function of QPO frequency for each of the \obsnumber observations. In the range where the lags are zero or negative ($\nu \gtrsim 1.8$~Hz), we find moderate values for $\delta H_{\rm ext} \approx 10$--20\%, decreasing to $\la$10\% for $\nu \ga 4$~Hz. Similar values are found for frequencies $\nu \la 1$~Hz, when the lags are positive. On the other hand, in the intermediate frequency range of 1--1.8~Hz, we find large values of this parameter, even above 100\%. This effect was also found by \cite{2021MNRAS.503.5522K}. In our spectral-timing model, $\delta H_{\rm ext}$ works as a normalisation of the fractional rms amplitude spectrum (see Sec.~\ref{sec:model}). This parameter has a dependence with the assumed size of the soft-photon source. In their paper, \cite{2021MNRAS.503.5522K} took this as a fixed parameter at 10~km, and make predictions of the external heating rate. In this paper, we assumed a larger value, equal to the typical inner radius of the accretion disk ($\sim$250~km at $\nu_0 = 2$~Hz, for a 12~M$_\odot$ BH in the LTP model). Considering a larger emitting area could, in principle, decrease those values, but a full new set of table models would be required. Given the fact that this parameter is non-sensitive to the lag spectrum, and that it only works as a normalisation, in that case, the rest of the parameters would remain essentially the same, which makes not worth to re-build all the table models and re-fit the data again. 

\begin{figure}
 \centering
 \includegraphics[width=\columnwidth]{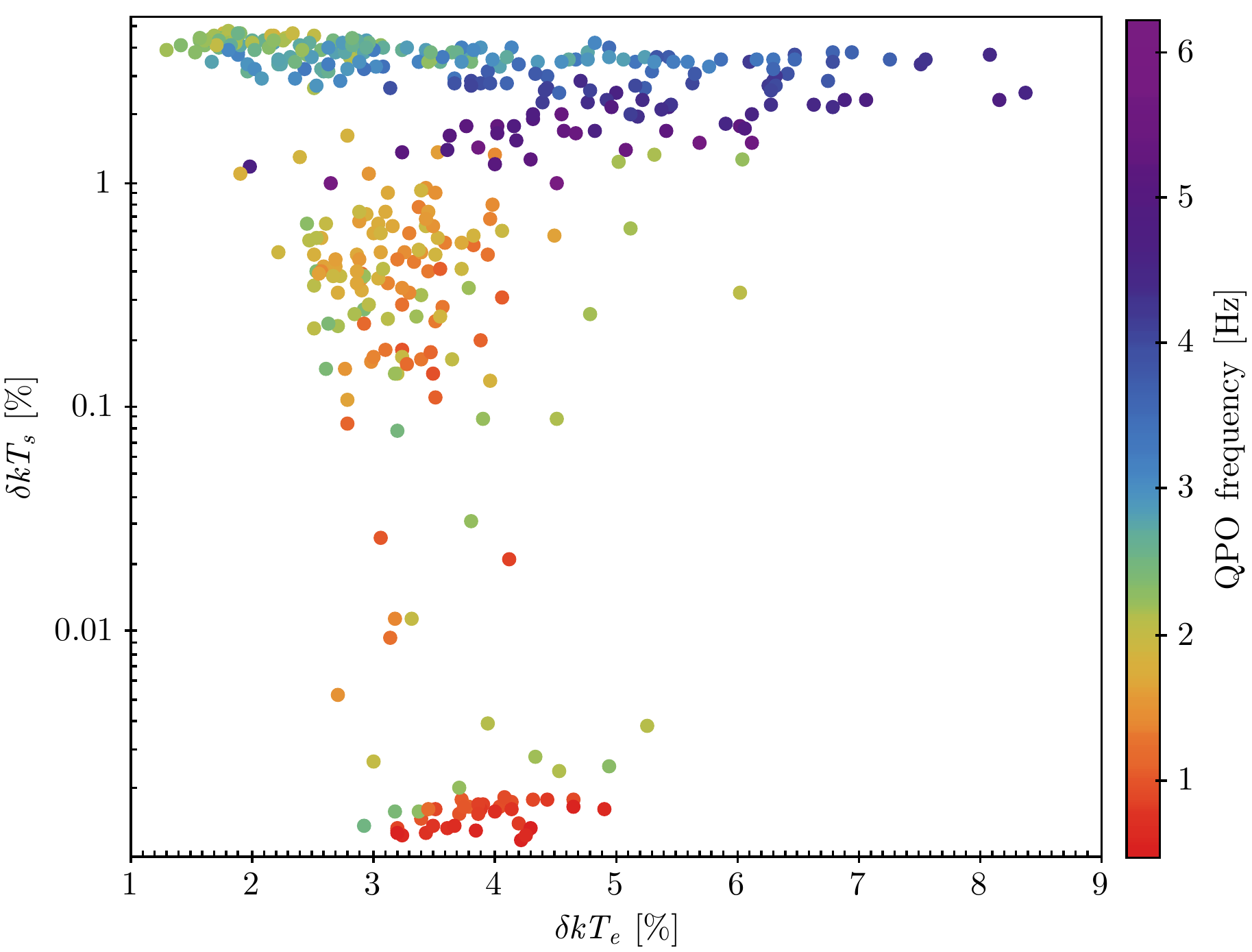}
 \caption{Amplitudes of the variability of the electron temperature in the corona, $\delta kT_e$, and of the temperature of the soft-photon source, $\delta kT_s$, in percent units. Values are always below 10\% in accordance with the linearised scheme considered in the spectral-timing Comptonisation model. The points are coloured according to the QPO frequency. }
 \label{fig:dTs}
\end{figure}

In order to verify if the solutions still remain consistent with the linearised regime assumed to calculate the solution of the Kompaneets equation in the spectral-timing model \citep[see][]{2020MNRAS.492.1399K}, in Fig.~\ref{fig:dTs} we present the fractional amplitudes of the variability of both the electron temperature in the corona, $\delta kT_e$, and the temperature of the soft-photon source, $\delta kT_s$, in percentage units. As seen in the plot, both quantities remain below 10\% for every observation, which is in accordance with the linearised approach used. In this Figure, we colour the points according to the QPO frequency to make evident the different regimes. In particular, it is remarkable that while $\delta kT_e$ is constrained to amplitudes of $\sim$1--9\%, $\delta kT_s$ spans for a much larger range, depending on the QPO frequency (and, hence, on the sign of the lags). For QPO frequencies $\nu \la 2.5$~Hz, $\delta kT_s < 1$\%, and, in particular, for $\nu \la 1.5$~Hz, when the lags become hard and the temperature of the soft photon source becomes very low (and a disc black-body in the time-averaged spectra is only given as an upper-limit), the amplitude of the oscillation of the temperature of the soft-photon source becomes very small ($\delta kT_s \la 100 \, \delta kT_e$), possibly indicating a change in the soft-photon source nature (see Discussion in Sec.~\ref{sec:discussion} above).

\section{Goodness of fit of the spectral-timing model}
\label{app:goodness}

\begin{figure}
 \centering
 \includegraphics[width=\columnwidth]{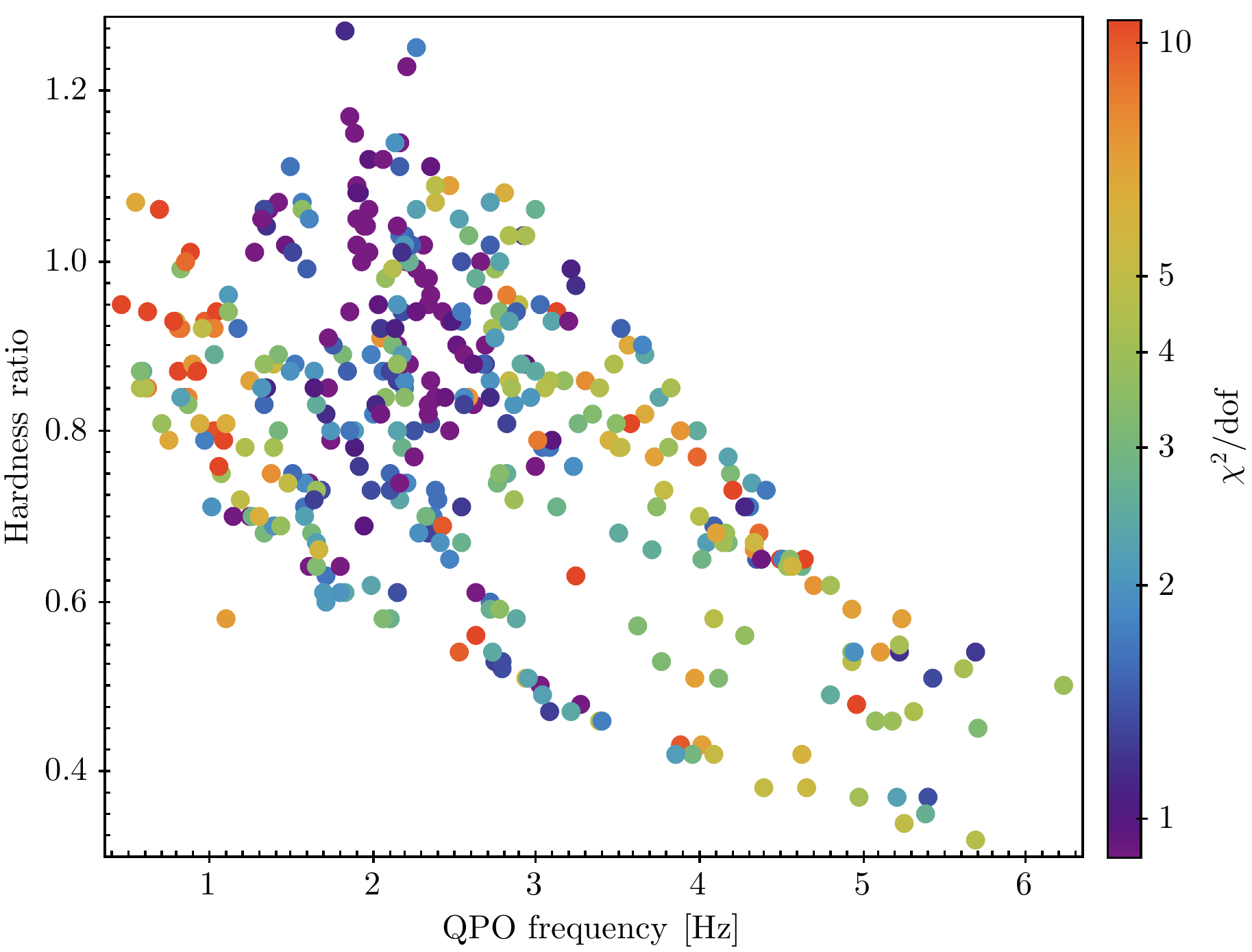}\\
 \includegraphics[width=0.9\columnwidth]{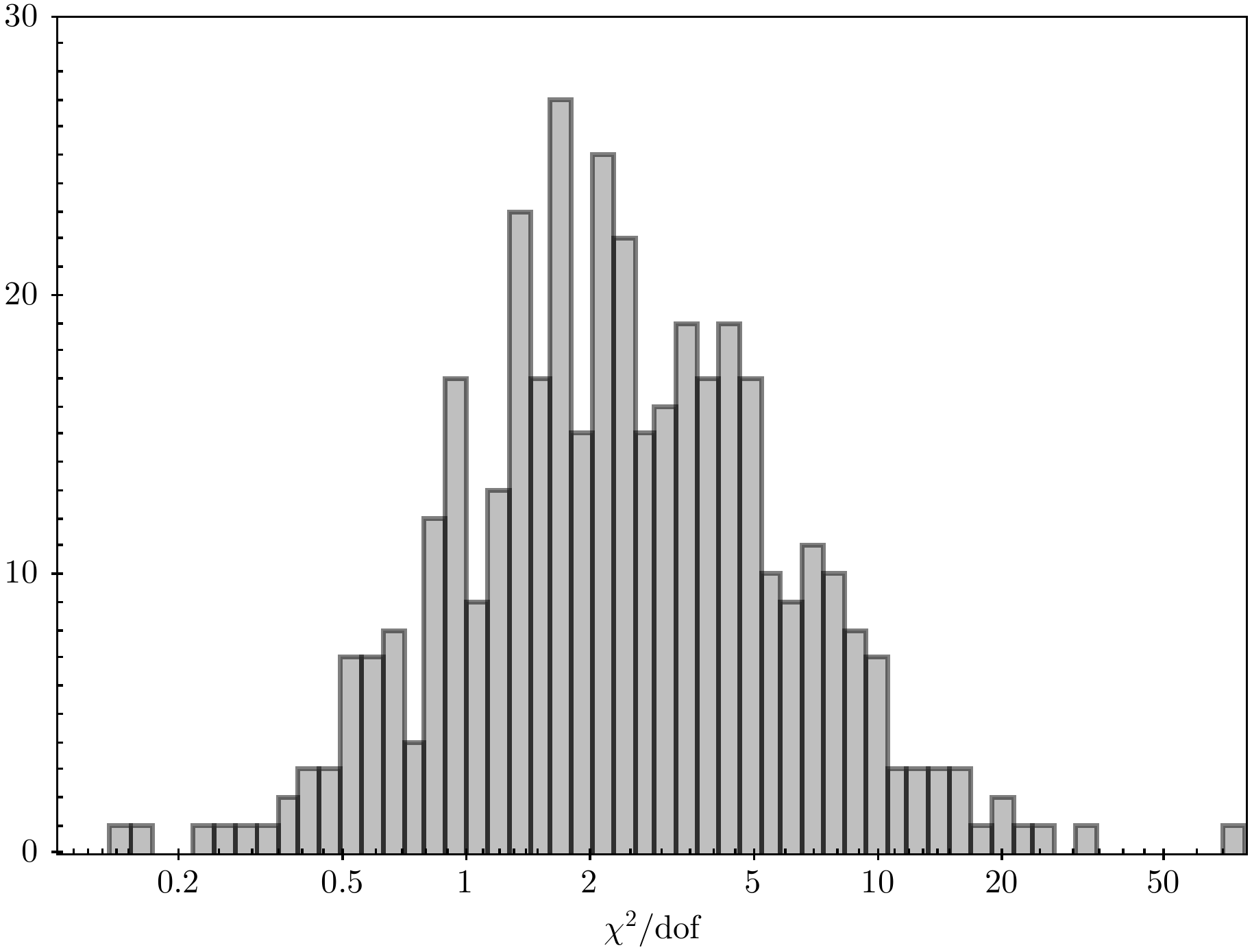} 
 \caption{Reduced chi-squared residuals, $\chi^2$/dof, of the best-fitting models obtained for the \obsnumber observations, as function of QPO frequency and hardness ratio (top panel) and as a marginalised distribution (histogram, bottom panel).}
 \label{fig:redchi}
\end{figure}

In Fig.~\ref{fig:redchi} we present the residuals obtained for the best-fitting models corresponding to each of the \obsnumber observations considered in this work. On the top panel we show a $QPO-HR$ diagram of the reduced $\chi$-squared residuals ($\chi^2$/dof). The model can fit the data well in the full frequency range. In particular, data is very-well fitted in the 1.0-3.5~Hz frequency range (with $\chi^2$/dof < 2) while less accurate fits are found for observations with either lower or higher QPO frequencies. On the bottom panel, we show a histogram corresponding to the distribution of reduced $\chi$-squared residuals obtained from the best-fitting models of the full set of observations. The distribution spans from 0.1 to $\sim$50, peaking in the 1--2 range, as expected for good fits.


\bsp	
\label{lastpage}
\end{document}